\documentclass[12pt]{iopart}
\usepackage{iopams, graphicx}

\DeclareMathAlphabet{\openchar}{U}{bbold}{m}{n}
\newcommand{\id}{\openchar{1}}

\renewcommand{\vec}[1]{{\mathbf #1}}
\newcommand{\vechat}[1]{\hat{{\mathbf #1}}}

\begin{document}

\title{Symmetry indicators of band topology}

\author{Hoi Chun Po}
\address{Department of Physics, Massachusetts Institute of Technology, Cambridge, MA 02139, USA}
\ead{hcpo@mit.edu}
\vspace{10pt}
\begin{indented}
\item[]November 2019
\end{indented}

\begin{abstract}
Topological materials are quantum materials with nontrivial ground-state entanglement that are irremovable so long as certain rules, like invariance under symmetries and the existence of an energy gap, are respected. They  showcase unconventional properties like robust anomalous surface states and quantized physical responses. The intense research efforts in understanding topological materials result in a modernized perspective on the decades-old theory of symmetry representations in electronic band structures, and inspire the development of general theories that enable the efficient diagnosis of topological materials using only symmetry data. One example is the theory of symmetry indicators of band topology, which is the focus of this topical review. We will aim at providing a pedagogical introduction to the key concepts and constructions in the theory, alongside with a brief summary of the latest development.
\end{abstract}

%
%
%
%
%

\section{Introduction}
Topological phases of matter are characterized by robust quantum entanglement in its ground state wave function. 
While typical examples of intrinsic topological orders \cite{Wen1995}, like the fractional quantum Hall states, only arise in interacting quantum systems, the integer quantum Hall states could be described within the approximation of non-interacting electrons \cite{TKNN}. Haldane's theoretical proposal for realizing the quantized Hall conductance without relying on an external magnetic field \cite{Haldane} demonstrates the possibility that quantum materials could be inherently topological. However, the relevance of this proposal to any realistic material system remained elusive for the decade which ensued.

The prospect for discovering realistic topological materials changed dramatically with the advent of topological insulators (TIs) \cite{RMP_TI, RevModPhys.83.1057, Molenkamp, BernevigBook}, which, unlike phases with intrinsic topological orders, require symmetry protection for their nontrivial ground-state quantum entanglement. 
In fact, the discovery of TIs shed light on a route for experimentally realizing the quantum anomalous Hall effect \cite{Chang167, QAH_Qi, QAH_Xue}, although the latter does not require symmetry protection.
While the original proposals focused on two-dimensional systems with spin-orbit coupling and  time-reversal symmetry \cite{Z2_QSH, Z2_Graphene,PhysRevLett.97.236601,Bernevig1757}, it was soon realized that the concept of symmetry-protected topological order \cite{Gu_Wen, PhysRevB.81.064439, PhysRevB.83.035107} could be extended to a plethora of symmetry settings
\cite{
PhysRevLett.98.106803, Joel_Z2, PhysRevB.79.195322, RMP_TI, RevModPhys.83.1057, Molenkamp, BernevigBook, Senthil_SPT}. Remarkably, many of the nontrivial fermionic phases can be realized even in systems of non-interacting fermions, and the notion of Chern numbers has been generalized to a variety of symmetry-protected topological invariants. 
This is exemplified by the richness displayed in the class of topological crystalline insulators (TCIs) \cite{PhysRevLett.106.106802, Slager, Hsieh, TCI_review, PhysRevB.93.045429, Wang:2016aa, Benalcazar61,PhysRevB.96.245115,PhysRevLett.119.246402,PhysRevLett.119.246401, Wieder246,Schindlereaat0346, PhysRevB.97.205136}, which reflects the complexity of the spatial symmetries in a crystal. The notion of topological materials has also been extended to semimetallic systems \cite{RevModPhys.90.015001}, like Weyl \cite{PhysRevB.83.205101, PhysRevX.5.011029, Huang:2015aa} and Dirac \cite{PhysRevLett.108.140405, PhysRevB.85.195320, PhysRevB.88.125427} semimetals, where the low-energy degrees of freedom arise from topologically protected nodal gap closing showcasing varied forms of dispersion, dimensionality, connectivity, and linking \cite{PhysRevB.84.235126, PhysRevB.92.081201, Soluyanov:2015aa, Bzdusek:2016aa,Bradlynaaf5037, PhysRevLett.121.106403}. Note that, in the current review, we will limit the scope of topological materials to those that could be modeled as weakly correlated electronic systems.

Among the many aspects of the extensive field of topological materials, two theoretical questions are particularly pertinent to the current review. The first concerns the problem of discovering new symmetry-protected topological phases, and subsequently classifying them, as was achieved in the ten-fold way classification for internal symmetries \cite{PhysRevB.55.1142, Kitaev, PhysRevB.78.195125}.
The mathematical concept of K theory has proven to be a very powerful classification framework
\cite{PhysRevLett.95.016405, Kitaev, Freed2013, PhysRevB.95.235425, PhysRevX.7.041069, shiozaki2018atiyahhirzebruch}. However, when the richness of spatial symmetries is taken into account, a concrete computation of the K-theoretic classification becomes highly technical and challenging. Furthermore, the physical interpretation of the classification is not always transparent. It would be desirable to interface these ideas with concepts that are more directly tied to the physical properties of a band structure.

The second question concerns the diagnosis and prediction of realistic topological materials. Generally, topological band invariants are defined using the global properties of the Bloch wave functions of the filled bands over the entire Brillouin zone. A direct numerical evaluation of such invariants in a realistic modeling of materials, as in {\it ab initio} methods, is computationally expensive in general. While clever workarounds have been advanced for important cases like in the computation of the Chern number and the $\mathbb Z_2$ TI indices \cite{PhysRevLett.107.036601,PhysRevB.89.115102}, one has to revisit and possibly revise the methods whenever a new topological (crystalline) invariant is proposed.

At the core of the study of topological materials is the interplay between symmetry and topology in the electronic band theory. Symmetry representation has always been a central subject in the development of band theory; what could we learn from this well-established subject? The Fu-Kane parity criterion \cite{PhysRevB.76.045302} provides an important lesson: in materials with inversion symmetry, the computation of the $\mathbb Z_2$ indices can be reduced to simple combinations of the inversion eigenvalues of the filled bands. This provides an extremely efficient shortcut compared to the evaluation of the wave-function based invariant, and has played a key role in the identification of material candidates for TIs \cite{RMP_TI, RevModPhys.83.1057, Molenkamp, BernevigBook}. Similar symmetry-based diagnosis for topological materials has also been proposed for other invariants, like the partial detection of Chern number using rotation eigenvalues \cite{PhysRevB.83.245132,Ari,PhysRevB.86.115112}.

The theory of symmetry indicators of band topology \footnote{It was originally introduced with the name ``symmetry-based indicators of band topology,'' but subsequent works tend to simplify the name by dropping the word ``based.'' We will adopt the simplified terminology here.} can be viewed as the systematic generalization of the Fu-Kane parity criterion to all symmetry settings and forms of topological materials \cite{Po2017}. It was motivated by recognizing that the modern topological point of view can significantly simplify the symmetry analysis in band theory, which then enables an efficient filtering of the topological band structures from the trivial ones. 
Since the symmetry-based methods \cite{Po2017, TQC}  only utilize the symmetry representations at isolated high-symmetry momenta as input, they enable the efficient diagnosis of topological materials, as is reflected in the large-scale topological materials discoveries which have been recently achieved 
\cite{Chen:2017aa, Tang2019_NP, Zhang2019, Vergniory2019, Tang2019, Tangeaau8725}. In addition, the mathematical structure of the symmetry indicators is closely related to the K theoretic classifications of TIs and TCIs. The computation of the symmetry indicators \cite{Po2017, Watanabeeaat8685}, together with a systematic analysis of their physical meaning \cite{PhysRevX.8.031069, PhysRevX.8.031070, QuantitativeMappings}, has enabled progresses in overcoming the technical difficulties in the classification problem \cite{shiozaki2018atiyahhirzebruch, PhysRevX.8.031070, Song2018, shiozaki2019classification}

This topical review aims at providing a pedagogical introduction to the theory of symmetry indicators, where the key ideas are either motivated or illustrated using concrete, worked examples. As a compromise, certain results will be stated without proof, and we will only sketch some of the required modifications for handling the general cases. 
This review could complement the existing works in the literature, which develop the theory with full generality at the cost of adopting a more abstract narration. We will also summarize some of the latest developments in areas related to the theory of symmetry indicators.

\section{Overview \label{sec:overview}}
Topological materials are characterized by symmetry-protected entanglement in the electronic ground-state wave function. In contrast, the familiar categories of insulating compounds well-described by either ionic or covalent bonding between the elements are topologically trivial, since their ground-state wave functions can be smoothly deformed to a product state described by the tight localization of the electrons to either  atomic or inter-atomic orbitals. We refer to such classes of trivial phases of matter as ``atomic insulators.'' By definition, it is impossible to smoothly deform between two topologically distinct phases of matter. For instance, one could imagine starting with a TI protected by time-reversal symmetry and gradually reduce the electron hopping amplitude until, say, an ionic insulator becomes the ground state. So long as the time-reversal symmetry is preserved throughout the deformation process, the system must go through a phase transition at which the energy gap closes.

The discussion above is general and is applicable even in the presence of strong electron-electron interactions and/ or disorder. For the purpose of the present review, however, we will specialize to systems of weakly correlated electrons in a crystal, which are well described by electronic band theory. More technically, such systems are descried by Hamiltonians which are quadratic in the fermion operators, preserves the total fermion number in the system, and possesses lattice translation symmetry.
The main simplification coming from this specialization is that the properties of the system are fully encoded in the Bloch Hamiltonian $H(\vec k)$, where the momentum $\vec k\in T^d$, the $d$-dimensional Brillouin zone. 
Now, suppose we have two systems with the same set of symmetries described respectively by the Bloch Hamiltonians $H_0(\vec k)$ and $H_1(\vec k)$, and suppose that there is a continuous energy gap between the $\nu$-th and ($\nu+1$)-th energy eigenstates for each of  $H_{0,1}(\vec k)$.
We can then conclude the two systems belong to the same phase of matter if one can find a smooth family of Hamiltonians $\{ H_{t}(\vec k)~:~ t \in [0,1]\}$ interpolating between the two while maintaining all the symmetries and the gap condition. 

Such a definition, however, is sometimes too restrictive on physical grounds, since certain deformation obstruction might be resolvable if one append a collection of trivial degrees of freedom to both $H_{0}$ and $H_1$, and when that is the case we say the two systems are ``stably equivalent.''
The Hopf insulator  \cite{PhysRevLett.101.186805, PhysRevB.88.201105,PhysRevB.95.161116} provides one such example, since its topological nature relies on the existence of maps that are relevant only when the system has exactly two bands. As such, the Hopf insulator is expected to be trivialized when additional trivial bands are introduced into the system \cite{PhysRevB.88.201105,PhysRevB.95.161116}.
Furthermore, the definition above does not elucidate on the mathematical structure between the different phases within the same symmetry setting. K theory has been introduced as a powerful mathematical framework for tackling these problems  \cite{PhysRevLett.95.016405, Kitaev, Freed2013, PhysRevB.95.235425, PhysRevX.7.041069, shiozaki2018atiyahhirzebruch}.
A key lesson from the K-theory framework is that topological distinctions between phases of matter with the same symmetries can be identified with the elements of an abelian group.
\footnote{
To elaborate on this point, we note that while it is natural to define addition of vector spaces (and correspondingly, vector bundles) by the direct sum, it is not obvious what would serve as an additive inverse of a given vector space. While one might then consider the space of band insulators as a commutative monoid due to the absence of inverses, an important insight from K theory is that one could instead define a suitable group which captures important aspects of the homotopic distinctions through the Grothendieck completion. Loosely, the idea is to introduce a comparison scheme by focusing on the formal difference between a pair of objects, and the inverse of any pair can then be obtained by simply reversing the order. Readers who are interested in a more in-depth discussion of these ideas are encouraged to consult the references listed above. In particular, we highlight that Appendix B in Ref.\ \cite{PhysRevX.7.041069} provides a concise introduction.
}

While the K-theoretic classification of weakly interacting electronic phases of matter is conceptually elegant and powerful, there are three main difficulties in directly utilizing the results for the discovery and diagnosis of topological materials. First, the computation is technically challenging when spatial symmetries are incorporated, and sophisticated mathematical methods and constructions are needed to arrive at a concrete classification.
Second, even when a topological invariant could be explicitly constructed following the abstract K-theoretic classification, the computation of the invariant generally requires knowledge of the Bloch wave functions over the entire Brillouin zone (numerically, a sufficiently dense mesh). This makes the computation of these invariants challenging in a realistic modeling of materials, say within the density functional theory. Lastly, in the typical formulation of K-theoretic classification, one is ultimately concerned with the topological {\it distinctions} between classes of Bloch Hamiltonians, whereas the physical definition of a topological phase of matter, as stated in the beginning of this section, is tied to the existence of symmetry-protected quantum entanglement in the ground state wave function. These two notions of ``topology'' do not generally agree. More specifically, in the presence of spatial symmetries, it is quite common that one could construct two product-state wave functions (i.e., atomic insulators) which cannot be smoothly deformed into each other while maintaining all symmetries (and allowing for the addition of ancilla degrees of freedom), i.e., the two phases are topologically distinct although neither of them is topological. Furthermore, an abrupt (i.e., not smooth) symmetric interface between these two phases may not harbor anomalous surface states.

The method of symmetry indicators for band topology is devised to fill the gap between the mathematical classification and the practical diagnosis of topological materials. Loosely, it could be viewed as a ``snapshot'' of the K-theoretic classification, in which, instead of discussing the topological distinctions between two collections of Bloch states defined by an energy gap, one focuses only on the symmetry representation furnished by the Bloch states at the high-symmetry points. Clearly, only partial information about the system is used, and so, generally, one does not obtain the full classification. This simplification, however, allows one to immediately conclude that, given a space group, any band structure compatible with an energy gap at all high-symmetry momenta can be viewed as an element of the abelian group $ \mathbb Z^{d_{\rm BS}}$. 
Here, the integer $d_{\rm BS} >0$ could be viewed as the dimension of the ``space'' of band structure, similar to the dimension of a vector space.
We will denote this group as $\{ {\rm BS}\}$, where BS stands for ``band structures,'' defined as the collection of Bloch-state symmetry representations which could be consistently patched together while maintaining an energy gap over all the high-symmetry momenta \cite{Po2017}.  
Such consistency requirements are known as ``compatibility conditions'' in band theory. 
Unlike the general computation of topological band invariants using the Bloch wave functions over a sufficiently dense mesh of the Brillouin zone, to diagnose a material from this perspective one only needs knowledge of the wave functions at isolated, high-symmetry momenta. 
Note our use of the phrase ``band structure'' but not ``band insulator,'' which reflects the fact that we are only imposing a gap condition over the high-symmetry momenta. This is a weaker condition than imposing it over the entire Brillouin zone, the latter of which is required for getting a band insulator.
In addition, in our interpretation the elements of $\{ {\rm BS}\}$ describe classes of Hamiltonian, instead of the distinctions between classes.
We will provide concrete examples for the notion of compatibility relations in \sref{sec:band_review} and the computation of the group $\{ {\rm BS}\}$ in \sref{sec:BS}.

Once a symmetry setting is specified, one can compute $\{ {\rm BS} \}$ directly by a systematic analysis of the compatibility relations in the momentum space. In particular, as an atomic insulator (AI) clearly sustains an energy gap and so satisfies all the compatibility relations, its momentum-space symmetry representations can be viewed as an element of $\{ {\rm BS}\}$. The physically interesting question is, does the consideration of all possible atomic insulators exhaust all the entries in  $\{ {\rm BS}\}$? If the answer is negative, there must be combinations of symmetry representations which are compatible with a band structure, but cannot arise from any atomic insulator. By definition, such band structures must be topologically nontrivial, and the nontriviality is diagnosable simply from analyzing the symmetry representations. The Fu-Kane parity criterion for TIs \cite{PhysRevB.76.045302} is an example of such diagnostics, and similar criteria have been previously discovered for the partial determination of the Chern number in the presence of a rotation symmetry \cite{PhysRevB.83.245132,Ari,PhysRevB.86.115112}.

The method of symmetry indicator aims to exhaustively expose all the stable forms of band topology which could be inferred from the symmetry representation data alone.
This is achieved by recognizing that the set of all possible atomic insulators automatically gives rise to a subgroup   $\{ {\rm AI}\}$ of $\{ {\rm BS}\}$. Abstractly, this subgroup again takes the form $\mathbb Z^{d_{\rm AI}}$, where the positive integer $d_{\rm AI} \leq d_{\rm BS}$. 
Since atomic insulators, by definition, admit a real-space description in which well-localized orbitals are fully filled, one could exhaustively tabulate all possible atomic insulators in a symmetry setting by combining all possible lattice types with the possible orbital types (defined in terms of symmetry representation).
The subgroup  $\{ {\rm AI}\} $ can then be identified by performing a Fourier transform to pass the  real-space symmetry data to the momentum space. Examples of this procedure are given in \sref{sec:AI}.

Once  $\{ {\rm AI}\}$ is known, nontrivial band topology can be exposed by analyzing the mismatch between $\{ {\rm AI}\}$ and $\{ {\rm BS}\}$. 
In other words, in order to expose topologically nontrivial materials it is desirable to forget about topological distinctions between atomic insulators. 
This is mathematically achieved by considering the quotient group 
\begin{equation}\label{eq:}
X_{\rm BS} \equiv \frac{\{ {\rm BS}\}}{\{ {\rm AI}\}},
\end{equation} 
which we will refer to as the ``(symmetry) indicator group.'' By construction, $X_{\rm BS}$ is always a finitely generated abelian group. From an explicit calculation, it was found that $X_{\rm BS}$ is a finite group for all of the 230 space groups, regardless of the presence or absence of time-reversal symmetry and spin-orbit coupling \cite{Po2017}. Equivalently, it was found that $d_{\rm AI} = d_{\rm BS}$ for all symmetry settings. Given this equality, all the compatibility relations can be inferred from one's knowledge of $\{ {\rm AI}\}$. As we will elaborate on later, one can find all the generators of $\{ {\rm BS}\}$ and compute $X_{\rm BS}$ once $\{ {\rm AI}\}$ is known.
Such computations will be illustrated in \sref{sec:SI}.

Before we move on to providing concrete examples on the mentioned ideas, we make a few remarks regarding the physical interpretation on the symmetry indicators. First, we reiterate that, given only the symmetry representations are utilized, one does not arrive at the full classification of topological (crystalline) insulators using the methods described here. This follows simply from the fact that, very often, the topology of a set of bands depends not only on the symmetry properties of the wave functions; any approach which  only utilizes symmetry data cannot be generally complete.

Second, we also emphasize again that our definition of a ``band structure'' only requires the compatibility with a continuous energy gap at all high-symmetry momenta, and, generally speaking, certain elements of $\{ {\rm BS}\}$ maybe necessarily gapless due to topological gap closings at some generic momenta. One example is the case of time-reversal breaking Weyl semimetals with a single pair of inversion-related Weyl points \cite{PhysRevB.83.245132,Ari}. This brings us to our third point: further analysis is required to understand the physical phases corresponding to a given symmetry indicator. Since the indicators do not  generally provide the full classification, certain distinct phases might collapse into the same indicator. Because of this, possessing a nontrivial symmetry indicator is only a sufficient, but not necessary, condition for a material to be topological. In other words, depending on the spatial symmetries present, certain topological phases may be missing from the symmetry-based diagnosis. Such is the case for TIs without any point-group symmetry, which cannot be diagnosed by symmetry indicators. The relations between symmetry indicators and topological phases of matter have been comprehensively studied for time-reversal-invariant materials with \cite{QuantitativeMappings, PhysRevX.8.031070} or without \cite{PhysRevX.8.031069} spin-orbit coupling, but the corresponding analysis for magnetic materials, which lacks time-reversal symmetry, remains open, although the symmetry indicator groups have already been exhaustively computed \cite{Watanabeeaat8685}.

\section{Review on symmetries of band structures \label{sec:band_review} }
In this review, we will be mostly interested in the action of spatial and time-reversal symmetries on electrons.
The symmetry representation theory associated with electronic energy bands is  well-established and many classic references are available, like \cite{Bradley}. In view of this, we will only provide a brief discussion here, with the goal of establishing notations, and also to highlight the main aspects of the representation theory which would be needed in developing the theory of symmetry indicators. We also remark that a more comprehensive review of the representation theory using a similar set of notations as here can be found in the Supplementary Note 1 of \cite{Po2017}.

\subsection{Representation of symmetries}
\begin{figure}[h]
\begin{center}
{\includegraphics[width=0.68 \textwidth]{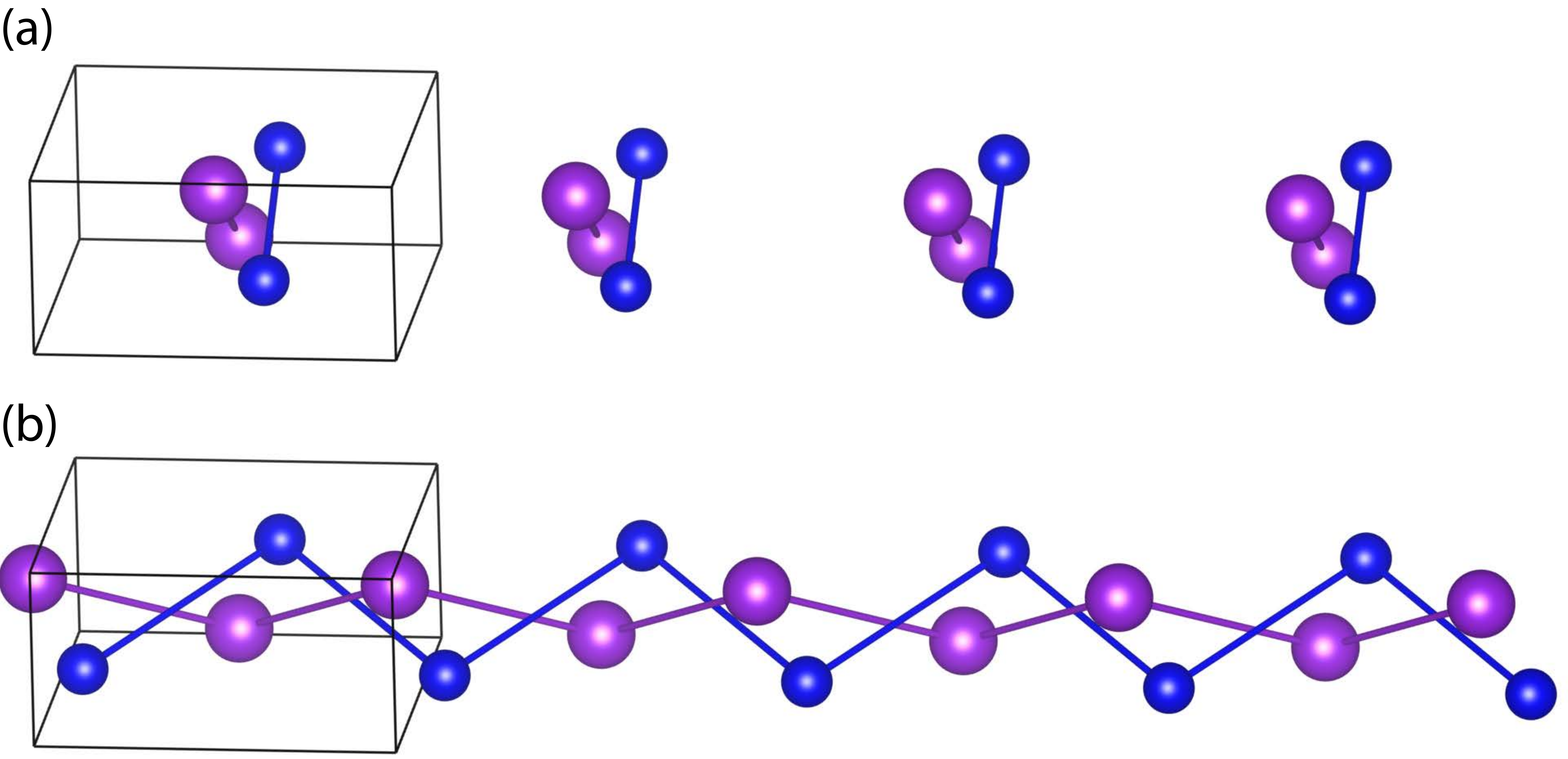}} 
\caption{ Examples of crystalline symmetries. (Figures are drawn with VESTA \cite{Vesta}.) (a) A two-fold rotation $C_{2x}$ about the center of the bonds maps the two purple (blue) sites in each unit cell to each other. Acting by $C_{2x}$ twice brings us back to the original site.
(b) A two-fold screw rotation $2_1$ combines a two-fold rotation with a half-lattice-translation. While it also maps the two purple (blue) sites to each other, acting by it twice brings us to a site in the next unit cell. 
\label{fig:C2_Screw}
 }
\end{center}
\end{figure}

Let $\mathcal G$ denote a three-dimensional space group. A general element $g \in \mathcal G$ acts on a point $\vec x \in \mathbb R^3$ by $g(\vec x) = p_g \vec x + \vec t_g$, where $p_g \in {\rm O}(3)$ is an element of the point group and $\vec t_g \in \mathbb R^3$. A common notation is to write $g = \{ p_g | \vec t_g \}$, and the multiplication of symmetry elements $g,h \in \mathcal G$, is given by 
\begin{equation}\label{eq:}
\{ p_g | \vec t_g \}\{ p_h | \vec t_h \} = \{ p_g p_h | \vec t_g  + p_g \vec t_h\}.
\end{equation} 
Note that $\vec t_g$ may not be a lattice vector. If there exists a choice of origin such that $\vec t_g$ is a lattice vector for all $g \in \mathcal G$, we say that $\mathcal G$ is {\it symmorphic}; otherwise we say it is {\it nonsymmorphic}. For most cases, one can view a nonsymmorphic symmetry as nontrivial ``roots'' of a lattice translation
\footnote{Technically, we are referring to ``intrinsically nonsymmorphic symmetries'' here. A nonsymmorphic space group may not contain such elements. Of the 230 three dimensional space groups, there are exactly two examples; see \cite{ITC, Konig:1999aa} for further discussions.}.
For instance, consider a $2_1$ screw symmetry along the $x$ direction (\fref{fig:C2_Screw}), given by combining the two-fold rotation $C_{2x}$ with a half-translation $\frac{1}{2}\vechat x $: $2_1 = \{ C_{2x} | \frac{1}{2} \vechat x\}$, where we have set the lattice constant to unity. One finds
\begin{equation}\label{eq:}
 \{ C_{2x} | \vechat x/2\}^2 = \{ E | \vechat x\},
\end{equation} 
the lattice translation along $x$ by one unit. Note that we denote the identity element of the point group by $E$.
In this sense, $2_1$ is a square root of $\{ E | \vechat x\}$.

When acting on momentum eigenstates like the Bloch states, the translation part of a spatial symmetry $g$ only gives rise to a phase, and the action of $g$ on the momentum $\vec k$ is given simply by the point-group action $g(\vec k) = p_g \vec k$.  We say $\vec k$ is left invariant by $g$ if $g(\vec k ) - \vec k = \vec G$ for some reciprocal lattice vector $\vec G$, and given a momentum $\vec k$, we define the little group $\mathcal G_{\vec k} \equiv \{ g\in \mathcal G ~:~ p_g \vec k = \vec k + ^\exists \vec G \}$.
We say $\vec k$ is a high-symmetry momentum whenever $\mathcal G_{\vec k}$ contains elements other then lattice translations.
For a crystal symmetric under the space group $\mathcal G$, the corresponding Bloch Hamiltonian $H_{\vec k}$ is symmetric under all elements  $ g \in \mathcal G_{\vec k}$, namely, there are unitary matrices $U_{\vec k}(g)$ such that
\begin{equation}\label{eq:}
U_{\vec k}(g)H_{\vec k} U^\dagger_{\vec k}(g) = H_{\vec k},
\end{equation} 
and the matrices satisfy 
\begin{equation}\label{eq:U_rep}
U_{\vec k}(g) U_{\vec k}(g') = z_{g,g'}U_{\vec k}(gg').
\end{equation} 
Here, $z_{g,g'} = \pm 1$ is a projective phase which is important when the system lacks spin-rotation invariance, say when spin-orbit coupling is significant 
\footnote{
In the crystallographic literature, instead of considering the projective representations, it is perhaps more common to consider an equivalent formulation in which one considers the non-projective representations of the ``doubled'' group. The doubled group is obtained basically by considering the fermion parity $(-1)^F$, typically denoted as $\bar E$, as an element of the group, and acknowledging that a $2\pi$ rotation gives $(-1)^F$ due to the spin-1/2 nature of the electrons, e.g., $(C_{2x})^2 = \bar E$. The representations suitable for spinful electrons require $\bar E$ to be represented as $-\id$.
}. 
If the system is invariant under spin rotation, one could always combine the spatial symmetry, which technically also acts on the electron spin, with the opposite spin rotation. As a result, the electron is effectively spinless and the phase $z_{g,g'} = 1$ always, i.e., the representation of $\mathcal G_{\vec k}$ is non-projective. We will refer this case as ``spinless.'' In the literature, such representations are usually called ``single-valued,'' whereas the case taking into account the spin-1/2 nature of electrons is referred to as ``double-valued.''

As a concrete example, consider a space group with the point group $D_2 = \{ E, C_{2x}, C_{2y}, C_{2z}\}$, and for simplicity we set $\vec k = \vec 0$.  Note that the identity element $E$ will always be represented by the identity matrix $\id$.
All the symmetry elements of the group $D_2$ commute with each other, and when the electrons are effectively spinless, a possible two-dimensional representation is
\begin{equation}\label{eq:D2_spinless}
\eqalign{
U(C_{2x}) =& \left(
\begin{array}{cc}
1 & 0 \\
0 & -1
\end{array}
\right);\\
U(C_{2y}) =& \left(
\begin{array}{cc}
-1 & 0 \\
0 & 1
\end{array}
\right);\\
U(C_{2z}) =& \left(
\begin{array}{cc}
-1 & 0 \\
0 & -1
\end{array}
\right).
}
\end{equation} 
All the matrices are diagonal and they clearly commute. In particular, note that $U(C_{2x})U(C_{2y}) = U(C_{2x} C_{2y}) = U(C_{2z})$.
However, if spin-orbit coupling is significant, the representations should encode the projective phases coming from the spin-1/2 nature of electrons. A possible two-dimensional ``spinful'' representation is
\begin{equation}\label{eq:D2_spinful}
\eqalign{
\tilde U(C_{2x}) =& i \sigma_x =  \left(
\begin{array}{cc}
0 & i \\
i & 0
\end{array}
\right);\\
\tilde U(C_{2y}) =&  i \sigma_y = \left(
\begin{array}{cc}
0 & 1 \\
-1 & 0
\end{array}
\right);\\
\tilde U(C_{2z}) =& i \sigma_z =  \left(
\begin{array}{cc}
i & 0 \\
0 & -i
\end{array}
\right),
}
\end{equation} 
where $\sigma$ denotes the Pauli matrices. In this case, we have $\tilde U(C_{2x})^2 = \tilde U(C_{2y})^2 = \tilde U(C_{2z})^2 = - \sigma_0$, and $\tilde U(C_{2x})\tilde U(C_{2y}) = - \tilde U(C_{2z})$.

Let us make a clarificatory remark: it is often stated that nonsymmorphic symmetries can also lead to projective representations, but this sense of projective representation is different form the one above. In the above, we consider the little group $\mathcal G_{\vec k}$, which always contains the subgroup $T$ of all lattice translations. The representation of a lattice translation $\{ E | \vec a\} \in T$ is $U_{\vec k}(\{ E | \vec a\}) = e^{- i \vec k \cdot \vec a} \id$, where $\id$ denotes the identity matrix of the same dimension as the Bloch Hamiltonian.
As a computational trick, it is customary to relate the representation of $\mathcal G_{\vec k}$ to that of $\mathcal G_{\vec k}/T = \{ p_g ~:~ g \in \mathcal G_{\vec k}\}$, which is a subgroup of the full point group $\mathcal G/T$. In the crystallographic literature, $\mathcal G_{\vec k}/T$ is called the {\it little co-group}.
When $\vec k \neq 0$, the translation phases arising from nonsymmorphic symmetries generally imply  that the representation of $\mathcal G_{\vec k}$ is related to the projective representations of the group $\mathcal G_{\vec k}/T$, since the latter does not encode the translation factors in the group structure.
As a concrete example, consider again the $2_1$ screw $ \{ C_{2x} | \frac{1}{2} \vechat x\}$, and suppose that it generates the entire $\mathcal G$. Consider $\vec k = k_x \vechat x$, and the representation
\begin{equation}\label{eq:2_1_eig}
\left [ U_{\vec k}\left( \left \{ C_{2x} \left | \frac{1}{2} \vechat x \right. \right\} \right) \right]^2 =  U_{\vec k}( \{ E | \vechat x\})  = e^{- i k_x}.
\end{equation} 
The first equality implies the representation of $\mathcal G_{\vec k}$ is non-projective. However, as point group elements we have $C_{2x}^2 = E$, and so one could also interpret the phase $e^{-i k_x}$ as a projective phase in the representation of the two-element group $\mathcal G_{\vec k}/ T = \{ E, C_{2x}\}$.
Although relating the representations of $\mathcal G_{\vec k}$ to the projective representations of $\mathcal G_{\vec k}/T$ is a powerful computational trick, this trick has no special role to play in our formalism, since conceptually we will always consider the representations of $\mathcal G_{\vec k}$  directly. For this reason, the presence of nonsymmorphic symmetries does not lead to any additional difficulty in our formalism.

Now, suppose $|\psi_{\vec k} \rangle$ is an eigenstate of $H_{\vec k}$ with energy $E_{\vec k}$. One sees that $U_{\vec k}(g)|\psi_{\vec k} \rangle$, which may or may not be proportional to $|\psi_{\vec k}\rangle$, will also be an eigenstate with the same energy for all $g\in \mathcal G_{\vec k}$. More generally, suppose $| \psi_{i,\vec k} \rangle$ for $i=1,\dots, d$ is a set of degenerate eigenstates of $H_{\vec k}$. We have
\begin{equation}\label{eq:u_rep}
U_{\vec k}(g) |\psi_{j,\vec k} \rangle = |\psi_{i,\vec k} \rangle [u_{\vec k}(g)]_{i,j},
\end{equation} 
where $u_{\vec k}(g)$ is a $d$-dimensional unitary matrix. The collection of matrices $\{ u_{\vec k}(g) ~:~ g \in \mathcal G_{\vec k}\}$ satisfies the same multiplication rules as in equation \eref{eq:U_rep}, and so they furnish a $d$-dimensional representation of $\mathcal G_{\vec k}$. For a generic system, these $d$ eigenstates of $H_{\vec k}$ are degenerate only because they are all symmetry-related, which in our context is equivalent to saying the collection $u_{\vec k}(g)$ gives an irreducible representation (irrep) of $\mathcal G_{\vec k}$. 
Given any representation of $\mathcal G_{\vec k}$, one can always decompose it into sums of the irreps, and conversely one can construct any representation by considering direct sums of the irreps.
Given the space group $\mathcal G$, a chosen momentum $\vec k$, and the presence or absence of spin-rotation invariance, all the possible irreps have been exhaustively tabulated \cite{Bradley}, and could be retrieved from, for instance, the Bilbao Crystallographic Server \cite{Bilbao_rep,PhysRevE.96.023310}.

To illustrate these points, let us consider again the example of the $D_2$ point group. First focus on the spinless case, and assume the Bloch Hamiltonian is two-dimensional with the symmetries represented according to equation \eref{eq:D2_spinless}. A symmetric Hamiltonian takes the form $H_{\vec 0} \propto \sigma_z$, and let us focus on the non-degenerate eigenstate $| \psi_{\vec 0} \rangle = (0,1)^T$. One finds
\begin{equation}\label{eq:}
u(C_{2x}) = -1;~~
u(C_{2y}) = 1;~~
u(C_{2z}) = -1,
\end{equation}
which is a particular one-dimensional irrep of $D_2$. 
In other words, the representation in equation \eref{eq:D2_spinless} is reducible, as is apparent from its diagonal form.
In contrast, for the spinful case, demanding invariance under conjugation by all the unitary matrices in equation \eref{eq:D2_spinful} leads to $\tilde H_{\vec 0} \propto \sigma_0$, i.e., the two states are degenerate. This is consistent with the observation that the representation in equation \eref{eq:D2_spinful} is already irreducible.

So far, we have only focused on the unitary spatial symmetries. The presence of time-reversal symmetry can lead to additional constraints on the symmetry representations. Suppose $\vec k_0$ is a time-reversal invariant momentum (TRIM), i.e., $2 \vec k_0$ is a reciprocal lattice vector. The Bloch Hamiltonian satisfies 
\begin{equation}\label{eq:}
U(\mathcal T) H_{\vec k_0}^* U(\mathcal T)^\dagger = H_{\vec k_0},
\end{equation} 
for a unitary matrix $U(\mathcal T)$. If $|\psi_{\vec k_0}\rangle$ is an eigenstate of $H_{\vec k_0}$, we see that 
$U(\mathcal T) | \psi^*_{\vec k_0}\rangle$ will also be an eigenstate with the same energy.
As time-reversal and spatial symmetries always commute, we have
\begin{equation}\label{eq:}
U(\mathcal T) U^*_{\vec k_0}(g) U(\mathcal T)^\dagger =U_{\vec k_0}(g)
\end{equation} 
for $g \in \mathcal G_{\vec k_0}$. Combining with equation \eref{eq:u_rep}, one sees that the states $\{ U(\mathcal T) | \psi^*_{i, \vec k_0}\rangle \}_{i=1}^d$ transforms under $\mathcal G_{\vec k_0}$ according to the conjugated representation $u_{\vec k_0}^*(g)$. 
One should ask if the two sets of states $\{  | \psi_{i, \vec k_0}\rangle \}_{i=1}^d$ and  $\{ U(\mathcal T) | \psi^*_{i, \vec k_0}\rangle \}_{i=1}^d$ are distinct or not. 
If $u_{\vec k_0}(g)$ and $u_{\vec k_0}^*(g)$ are different irreps of $\mathcal G_{\vec k_0}$, the two sets must be distinct and all these $2d$ states are degenerate. However, even if the two sets give rise to the same irrep of $\mathcal G_{\vec k_0}$, they may still be orthogonal.  
To summarize, there are three possible outcomes when one considers the action of time-reversal symmetry on an irrep $\{ u_{\vec k_0}(g) : g \in \mathcal G_{\vec k_0}\}$: (i) the states furnishing the representation are closed under time-reversal, and so the degeneracy is unchanged; (ii) the irrep $u$ is paired with another copy of itself, doubling the degeneracy; or (iii) the irrep $u$ is paired with $u^*$, a different irrep, and so the degeneracy is doubled.
In practice, these different cases can be distinguished using the ``Wigner's test'' \cite{Bradley}, which relates the characters of the irrep to the three cases listed above.

Let us again illustrate the action of time-reversal symmetry using the $D_2$ example. In the spinless case, one could simply pick $U(\mathcal T) = \sigma_0$, and the two states $(1,0)^T$ and $(0,1)^T$, corresponding to two irreps, are individually time-reversal invariant, and so both of them fall into case (i) discussed above.
For the spinful case, a natural choice would be $\tilde U(\mathcal T) = i \sigma_y$, which satisfies $\tilde U(\mathcal T)  \tilde U^*(\mathcal T)  = - \sigma_0$. Here, the two states in the irrep form a Kramer's doublet under time-reversal symmetry, and so this is also an example of case (i). However, if we suppose that the $C_{2z}$ symmetry is broken while $C_{2x}$ is retained, the representation ceases to be irreducible in terms of the unitary spatial symmetries, as it splits into two irrep characterized by $u(C_{2x}) = \pm i$. Nevertheless, these two states still remain a Kramer's pair of time-reversal symmetry, and this gives an example of case (iii).

\subsection{Compatibility relations}
In the previous subsection, we have focused on the symmetry representations at a single high-symmetry momentum. As a band structure is defined globally over the Brillouin zone, it is important to understand how the symmetry representations at different momenta are related to each other. These relations are usually called ``compatibility relations.'' 
Let us consider a high-symmetry momentum $\vec k_0$ with the little group $\mathcal G_{\vec k_0}$. We consider moving away from $\vec k_0$ by a small distance $|\delta \vec k|$ such that some symmetries are broken at $\vec k_0 + \delta \vec k$. In other words, their little groups are related by $\mathcal G_{\vec k_0+\delta \vec k} \leq \mathcal G_{\vec k_0}$.
Suppose we are given a Bloch Hamiltonian $H_{\vec k_0}$, and that at $\vec k_0$ the states $\{ | \psi_{i,\vec k_0} \rangle\}_{i=1}^d$ furnish a $d$-dimensional irrep $u_{\vec k_0}$ of $\mathcal G_{\vec k_0}$. Generally, the degeneracy of these states are lifted once we go to $\vec k_0 + \delta \vec k$, since some of the symmetries relating the states are broken. Nevertheless, we can still consider the $d$-dimensional representation $U_{\vec k_0 + \delta \vec k}$ of $\mathcal G_{\vec k_0 + \delta \vec k}$ furnished by the (non-degenerate) states  $\{ | \psi_{i,\vec k_0+\delta \vec k} \rangle\}_{i=1}^d$.
As $|\delta \vec k|$ is infinitesimally small, the representation $U_{\vec k_0 + \delta \vec k}$ must be fully specified by $u_{\vec k_0}$: starting from $u_{\vec k_0}$, we simply restrict our attention to elements of $\mathcal G_{\vec k_0}$ which also belongs to $\mathcal G_{\vec k_0+\delta \vec k}$ to arrive at $U_{\vec k_0 + \delta \vec k}$. Technically, we say that $U_{\vec k_0 + \delta \vec k}$  is ``subduced'' from $u_{\vec k_0}$. Importantly, even if we start with an irreducible representation  $u_{\vec k_0}$ of $\mathcal G_{\vec k_0}$, the resulting representation $U_{\vec k_0 + \delta \vec k}$ of $\mathcal G_{\vec k_+\delta \vec k}$ is generally reducible. The representation $U_{\vec k_0 + \delta \vec k}$ can be specified by an expansion in terms of the irrep of $\mathcal G_{\vec k_+\delta \vec k}$, and the relations between $u_{\vec k_0}$ and $U_{\vec k_0 + \delta \vec k}$ are called compatibility relations.

\begin{center}
\begin{table}[h]
\caption{Irreducible representations of $D_2$ for spinless electrons
\label{tab:D2_spinless}}
\begin{tabular}{c|cccc}
~ & $E$ & $C_{2x}$ & $C_{2y}$ & $C_{2z}$\\
\hline
$A$ & $1$ & $1$ & $1$ & $1$\\
$B_1$ & $1$ & $-1$ & $-1$ & $1$\\
$B_2$ & $1$ & $-1$ & $1$ & $-1$\\
$B_3$ & $1$ & $1$ & $-1$ & $-1$
\end{tabular}
~\\
\caption{Irreducible representations of $C_2$ for spinless electrons
\label{tab:C2_spinless}}
\begin{tabular}{c|cc}
~ & $E$ & $C_{2}$\\
\hline
$A$ & $1$ & $1$\\
$B$ & $1$ & $-1$
\end{tabular}
\end{table}
\end{center}

To demonstrate these ideas concretely, let us revisit the example of $D_2$ point group. First, we consider the spinless case, for which the irreps are indicted in \tref{tab:D2_spinless}. Let us set $\vec k_0 = (0,0,0)$, the $\Gamma$ point, and consider $ \delta \vec k$ along the $(1,0,0)$ direction, bringing us to the $\Sigma$ line. Suppose $\mathcal G_{\Gamma}/T= D_2$. Along the line $\Sigma$, only $C_{2x}$ symmetry remains, so we have $\mathcal G_{\Sigma}/T = C_2$. By comparing \tref{tab:D2_spinless} and  \tref{tab:C2_spinless}, we can conclude the compatibility relations
\begin{equation}\label{eq:D2_CR_spinless}
\eqalign{
A &\mapsto  A;\\
B_1 &\mapsto  B;\\
B_2 &\mapsto  B;\\
B_3 &\mapsto  A.
}
\end{equation} 
For this particular example, the irreps are all one-dimensional, so there will not be any change in degeneracy when we move from $\Gamma$ to $\Sigma$. 

The story is different for the spinful case. Let us label the two-dimensional representation in equation \eref{eq:D2_spinless}  by $\bar E$. The two states, which have $C_{2x}$ eigenvalues $\pm i$, do not remain degenerate as we move to $\Sigma$. If we label these two irreps for $C_2$ by ${^{1}\bar E}$ and ${^{2}\bar E}$, then the compatibility relation can be written as
\begin{equation}\label{eq:D2_CR_spinful}
\bar E \mapsto  {^{1}\bar E} \oplus {^{2}\bar E}.
\end{equation} 
We remark that, in the literature, it is also a common practice to replace the irrep labels above by those specific for the momenta involved, e.g., equation \eref{eq:D2_CR_spinful} may be indicated as $\bar \Gamma_5 \mapsto \bar \Sigma_3 \oplus \bar \Sigma_4$ \cite{Bradley, Bilbao_rep, PhysRevE.96.023310}. 

\begin{figure}[h]
\begin{center}
{\includegraphics[width=0.48 \textwidth]{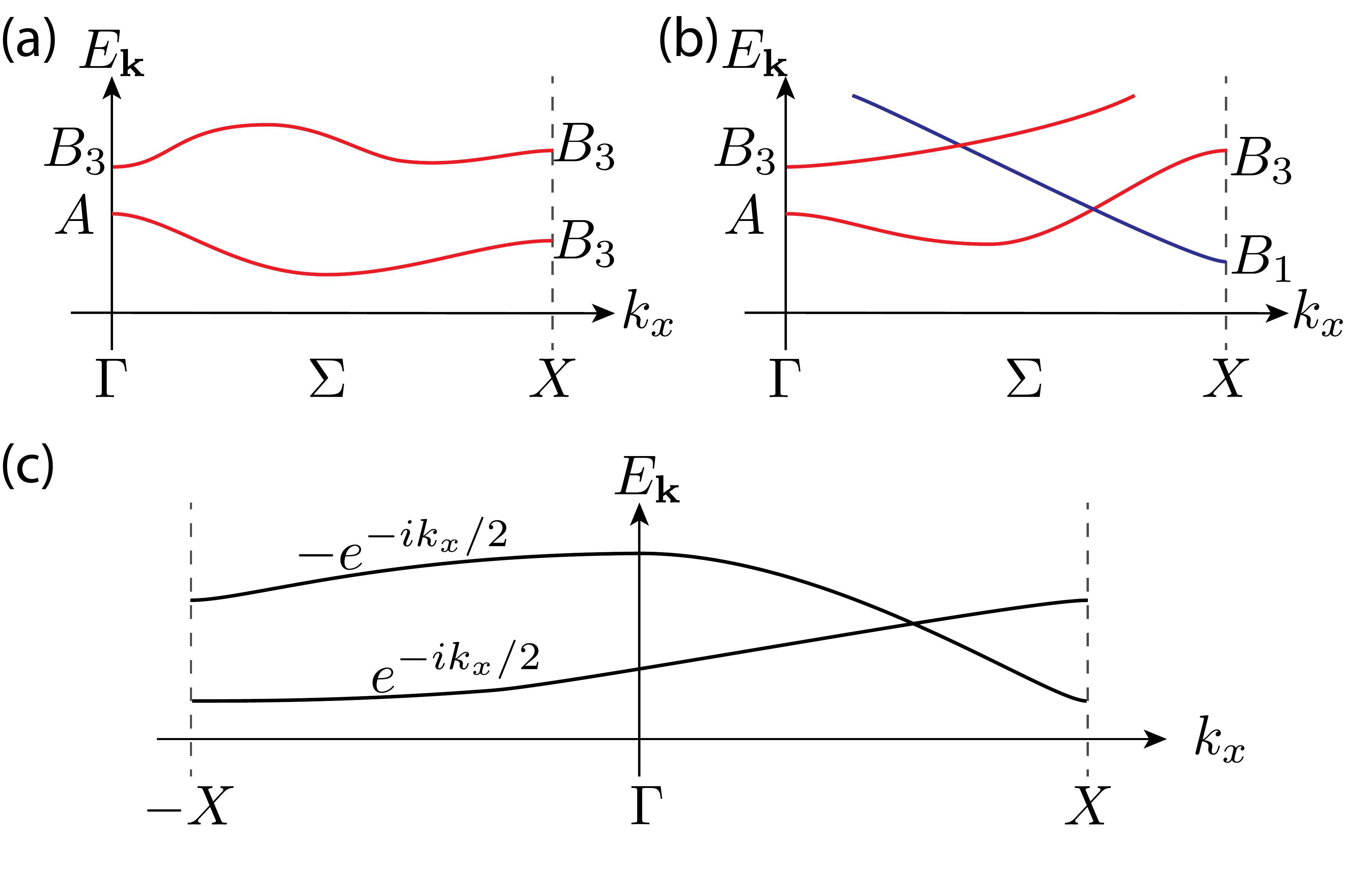}} 
\caption{Examples of compatibility relations. (a,b) Matching of representations between end points of a high-symmetry line. The color indicates $C_{2x}$ symmetry eigenvalues. (c) Enforced band connectivity due to nonsymmorphic symmetry.
\label{fig:CR}
 }
\end{center}
\end{figure}

The compatibility relations play a key role in determining if the energy bands can be gapped at all high-symmetry momenta. To illustrate the idea, suppose we have two energy bands of interest, and they furnish the irreps $A$ and $B_3$ of $\mathcal G_{\Gamma}/T = D_2$. 
Let us assume $\mathcal G_{X}/T = D_2$, where $X=(\pi,0,0)$. The two bands, however, can furnish different representations of $D_2$, say both of them transforming as $B_3$. 
We ask if these two bands can remain isolated from others along the line $\Sigma$. 
From equation \eref{eq:D2_CR_spinless}, we see that both $A\oplus B_3$ and $B_3\oplus B_3$ subduce to the representation $A \oplus A$ on $\Sigma$, and so the two end points are compatible and we conclude the two bands can remain isolated (\fref{fig:CR}a). Alternatively, suppose the two bands at $X$ realize the representation $B_1 \oplus B_3$. As $B_1\oplus B_3 \mapsto A\oplus B$, the two end points no longer give rise to the same representation on $\Sigma$. This implies there must be a gap closing between the two bands of interest and the rest, which fixes the difference in the representation content (\fref{fig:CR}b).

Nonsymmorphic symmetries can also lead to additional conditions on the symmetry representations of isolated bands. Let us again consider the $2_{1}$ screw $\{C_{2x}|\frac{1}{2} \vechat x\}$. As we can see from the spinless case in equation \eref{eq:2_1_eig}, a Bloch state along the line $(k_x,0,0) $ represents $\{C_{2x}|\frac{1}{2} \vechat x\}$ by $\pm e^{-i k_x /2}$. Suppose we start with a state $|\psi_{k_0} \rangle$ at $(k_0,0,0)$ with $2_1$ eigenvalue of $ e^{-i k_0 /2}$. If we follow this state smoothly as we traverse the Brillouin zone, the momentum changes by $k_0 \mapsto k_0 + 2\pi$, bringing us to the same point as we started, but the $2_1$ eigenvalue would have changed to $ e^{-i (k_0 + 2\pi) /2} = -e^{-i k_0 /2}$. In other words, there must have been a gap closing as we go from $k_0$ to $k_0 + 2\pi$ such that we end up at a different Bloch state (\fref{fig:CR}c). This implies the two representations $\pm e^{-i k_x/2}$ always come together in an isolated set of bands, which constitutes another class of constraints \cite{Michel_Zak_connectivity}. We will also refer to such nonsymmorphic constraints as compatibility relations, although this is not the typical terminology.

Lastly, we note that we have generally ignored the action of symmetries relating different points in the Brillouin zone. These symmetries will also cast constraints on the representations appearing at symmetry-related momenta. More explicitly, let $g \not \in \mathcal G_{\vec k}$, then the Bloch Hamiltonians $H_{\vec k}$ and $H_{g \vec k}$ will be symmetry-related, i.e., there is a unitary matrix $U_{\vec k}(g)$ such that $U_{\vec k}(g) H_{\vec k} U^\dagger_{\vec k}(g) = H_{g \vec k}$. Now, if a set of states $\{ | \psi_{i,\vec k}\rangle \}_{i=1}^d$ furnishes an irrep $u_{\vec k}$ of $\mathcal G_{\vec k}$, the states $\{ U_{\vec k}(g) | \psi_{i,\vec k}\rangle \}_{i=1}^d$ will furnish a corresponding irrep $u'_{g \vec k}$ of $\mathcal G_{g \vec k}$. One would also need to take these constraints into account in order to obtain a set of bands which are isolated at all high-symmetry momenta. By the same abuse of terminology, we will again refer to these constraints as part of the compatibility relations.
 
\subsection{Fourier transform
\label{sec:FT}}
Our discussion thus far focuses on the momentum space. In particular, we started with the general relation $U_{\vec k}(g) H_{\vec k}U^\dagger_{\vec k}(g) = H_{g \vec k}$ between the Bloch Hamiltonians at symmetry-related momenta, and discussed how, when focusing on $g \in \mathcal G_{\vec k}$, the possible representations $U_{\vec k}$ could be organized using the irreps of $\mathcal G_{\vec k}$. We have, however, never addressed how the original representation $U_{\vec k}$ arises in the first place. This subsection is devoted to filling this gap.

For simplicity, we will illustrate the ideas in the context of a tight-binding model, but similar discussion applies to other approaches for obtaining the Bloch Hamiltonian. Let us consider a general tight-binding Hamiltonian written in the real space basis:
\begin{equation}\label{eq:}
\hat H = \sum_{ i,j; \vec R, \delta \vec R} t_{i,j; \delta \vec R}\hat c^{\dagger}_{i, \vec R+\delta \vec R}  \hat c_{j, \vec R},
\end{equation} 
here $\vec R$ and $\delta \vec R $ are {\it unit cell} coordinates, and $i,j$ runs over all the degrees of freedom within a unit cell, like sites, orbitals and spins. We choose the Fourier transform convention
\begin{equation}\label{eq:c_FT}
\hat c^\dagger_{i, \vec k} \equiv \frac{1}{\sqrt{V}}\sum_{\vec R} \hat c^\dagger_{i, \vec R} e^{ - i \vec k \cdot \vec R},
\end{equation} 
where the sum on $\vec R$ runs over a system of volume $V$ assuming periodic boundary condition. The Bloch Hamiltonian is then given by
\begin{equation}\label{eq:}
[H_{\vec k}]_{ij} = \sum_{\delta \vec R}t_{i,j; \delta \vec R} e^{- i \vec k \cdot \delta \vec R}.
\end{equation} 
Notice that, in this convention, $H_{\vec k} = H_{\vec k + \vec G}$ for any reciprocal lattice vector $\vec G$. This is because we only Fourier transform with respect to the unit cell coordinates, and the fractional coordinates of the individual sites are all hidden in the site-orbital indices. An other common choice of Fourier transform would instead use the physical coordinates of each site, and in that case the Bloch Hamiltonians at $\vec k$ and $\vec k+ \vec G$ will be related by a $\vec G$-dependent unitary matrix. Our discussion would not depend on such details on the Fourier transform convention so long as the same convention is used throughout.

We are now ready to address the action of spatial symmetries. Generally, a symmetry element $g = \{p_g|\vec t_g\}\in \mathcal G$ transforms the fermion operators by 
\begin{equation}\label{eq:g_RS}
g: c^\dagger_{j, \vec R} \mapsto \sum_{i}  c^\dagger_{i, g_j(\vec R)}  U_{ij}(g),
\end{equation} 
where $U(g)$ is a unitary matrix encoding the transformation properties of the orbital associated with $c^\dagger_j$. Note the notation $g_j(\vec R)$: the action of $g$ on $\vec R$ is not as simple as $p_g \vec R + t_{g}$, which may not even be a lattice vector when $g$ is nonsymmorphic. 
Such complication arises when $g$ permutes the sites in a unit cell, and $g_j(\vec R)$ is defined to be the correct unit cell coordinate when such permutation is taken into account. 

In any case, given the real-space symmetry action in equation \eref{eq:g_RS}, we can plug it into to the Fourier transform in equation \eref{eq:c_FT}, and obtain the symmetry representation $U_{\vec k}(g)$ in the momentum space.
Although the idea is simple, and the computation for any specific case is also typically straightforward, discussing such calculations in full generality would unavoidably be quite technical. The general frameworks have already been reviewed in the supplementary materials of \cite{Po2017, TQC}, and also developed a bit more abstractly in the theory of {\it band representations} \cite{PhysRevLett.45.1025,PhysRevB.23.2824, PhysRevLett.61.1005, Bacry, MichelZak}. Here, we will refrain from reiterating such technical discussions, but rather attempt to illustrate the key ideas through a few simple examples.

\begin{figure}[h]
\begin{center}
{\includegraphics[width=0.68 \textwidth]{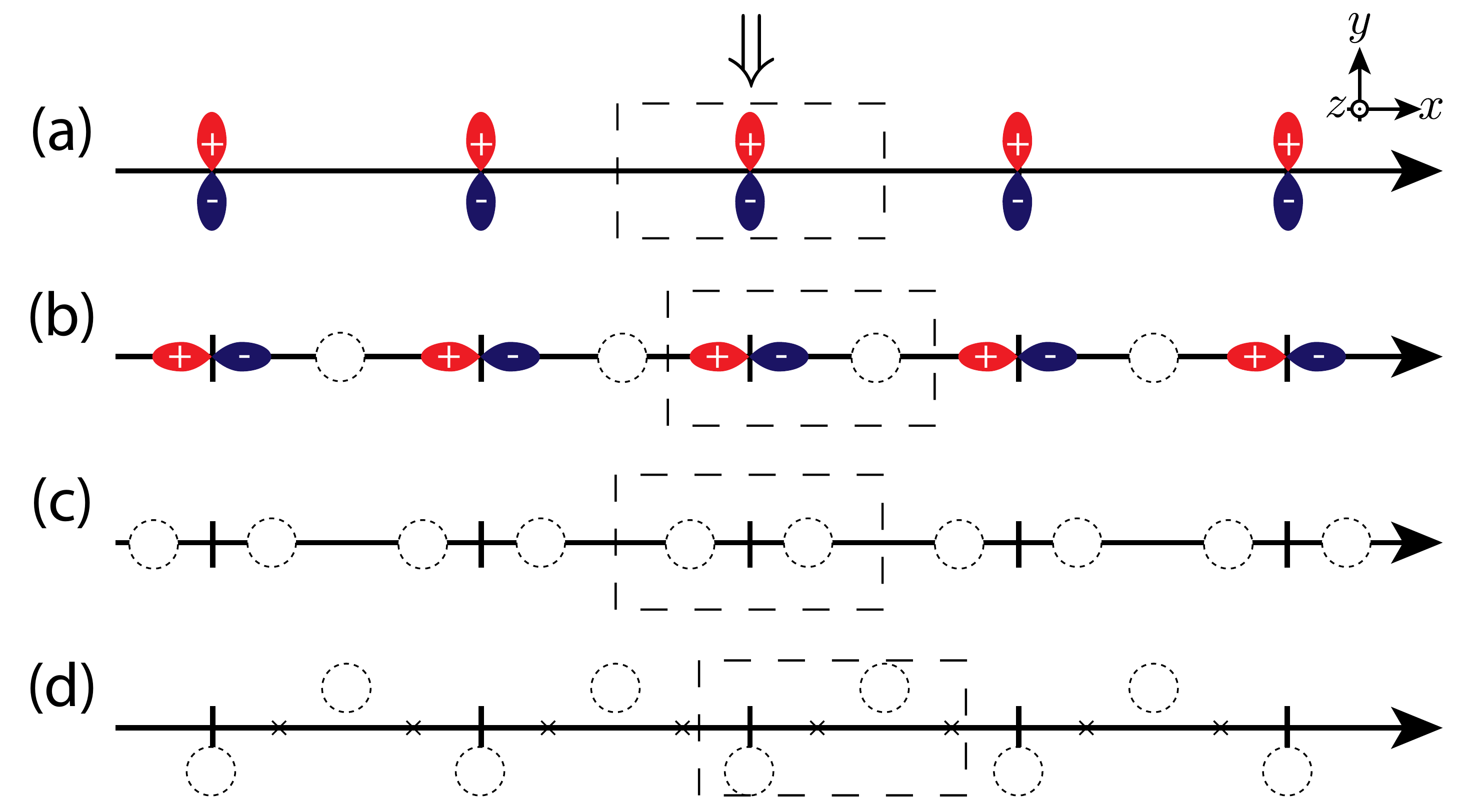}} 
\caption{Examples of symmetry representations from Fourier transforming real-space degrees of freedom. The double arrow indicates the origin, and orbitals assigned to the same unit cell are indicated by the boxes.
\label{fig:FT_ex}
 }
\end{center}
\end{figure}

Consider a one-dimensional chain oriented along the $x$ axis, and suppose that, in each unit cell, we have one $p_y$ orbital (\fref{fig:FT_ex}a). For simplicity, we will consider a spinless problem. We will again consider the $D_2$ point group, although, strictly speaking, our model has a higher degree of spatial symmetries (e.g., it also has inversion symmetry). 
In the following, we always consider the rotations about the origin, indicated in \fref{fig:FT_ex} by a double arrow.
The $p_y$ orbital picks up a $-1$ sign under the $C_{2x}$ rotation, but transforms trivially under  $C_{2y}$. In other words, it furnishes the $B_2$ representation in \tref{tab:D2_spinless}. In terms of the fermion operators, we have
\begin{equation}\label{eq:}
C_{2x}: \hat c_{x}^\dagger \mapsto -\hat c_{x}^\dagger; ~~~~~~
C_{2y}: \hat c_{x}^\dagger \mapsto \hat c_{-x}^\dagger,
\end{equation} 
where $x \in \mathbb Z$.
The symmetry properties of $\hat c_{\vec k}^\dagger$ then follows immediately:
\begin{equation}\label{eq:}
\eqalign{
C_{2x}:~& \hat c_{k_x}^\dagger \mapsto -\frac{1}{\sqrt{V}} \sum_{x} \hat c_{x}^\dagger e^{- i  k_x x} = -\hat c^\dagger_{k_x};\\
C_{2y}:~& \hat c_{k_x}^\dagger \mapsto \frac{1}{\sqrt{V}} \sum_{x} \hat c_{-x}^\dagger e^{- i k_x x} = \hat c^\dagger_{-k_x}.
}
\end{equation} 
In particular, we see that $c^\dagger_{k_x=0}$ and  $c^\dagger_{k_x=\pi}$ are invariant under the symmetries, and both of them furnish the irrep $B_2$ of $D_2$.

As our next example, consider the model in \fref{fig:FT_ex}b. At the origin, we have replaced the $p_y$ orbital by a $p_x$ orbital, and so it transforms as
\begin{equation}\label{eq:}
C_{2x}: \hat c_{1,x}^\dagger \mapsto \hat c_{1,x}^\dagger; ~~~~~~
C_{2y}: \hat c_{1,x}^\dagger \mapsto -\hat c_{1,-x}^\dagger.
\end{equation} 
By the same analysis as in the first example, we know that $\hat c_{1,k_x = 0,\pi}$ also furnishes the $B_3$ irrep.  
On top of that, we have also placed an $s$ orbital at $\tilde x =\pm \frac{1}{2}, \pm \frac{3}{2},\dots$. 
Since, in our convention, we only keep track of the unit-cell coordinate, we have to make a choice in assigning the half-integer sites to the unit cell, say let $\hat c_{2,x=0}$ denote the fermion localized to the site at $\tilde x=\frac{1}{2}$. With this choice, we see that
\begin{equation}\label{eq:half_int_sites}
C_{2x}: \hat c_{2,x}^\dagger \mapsto \hat c_{2,x}^\dagger; ~~~~~~
C_{2y}: \hat c_{2,x}^\dagger \mapsto \hat c_{2,-x-1}^\dagger.
\end{equation} 
Notice the action of $C_{2y}$ on the unit-cell coordinate: to see why that makes sense, note that the site at $\tilde x= \frac{3}{2}$ is assigned to the unit cell $x=1$, and that at $\tilde x = -\frac{3}{2}$ is assigned to the unit cell $x = -2$. With this preparation, we can consider the Fourier transform
\begin{equation}\label{eq:}
\eqalign{
C_{2x}:~& \hat c_{2, k_x}^\dagger \mapsto \frac{1}{\sqrt{V}} \sum_{x} \hat c_{2,x}^\dagger e^{- i  k_x x} = \hat c^\dagger_{k_x};\\
C_{2y}:~& \hat c_{2, k_x}^\dagger \mapsto \frac{1}{\sqrt{V}} \sum_{x} \hat c_{2,-x-1}^\dagger e^{- i k_x x} = \hat c^\dagger_{-k_x} e^{i k_x}.
}
\end{equation} 
Combining the analysis for the two sets of sites, we arrive at the representations
\begin{equation}\label{eq:}
\eqalign{
U_{\Gamma}(C_{2x}) =& \left(
\begin{array}{cc}
1 & 0 \\
0 & 1
\end{array}
\right),
U_{\Gamma}(C_{2y}) = \left(
\begin{array}{cc}
-1 & 0 \\
0 & 1
\end{array}
\right);\\
U_{X}(C_{2x}) =& \left(
\begin{array}{cc}
1 & 0 \\
0 & 1
\end{array}
\right),
U_{X}(C_{2y}) = \left(
\begin{array}{cc}
-1 & 0 \\
0 & -1
\end{array}
\right).
}
\end{equation} 
In other words, they furnish the representation $A\oplus B_{3}$ at $\Gamma$, but the representation $B_3\oplus B_3$ at $X$. Note that this is nothing but the representations shown in \fref{fig:CR}a, which, of course, are compatible with an isolated set of bands. In fact, $C_{2x}$ is a symmetry for all values of $k_x$, and we can simply write $U_{k_x}(C_{2x}) = \sigma_0$.

As our third example, let us now put two $s$ orbitals at $\tilde x = \pm 0.2, \pm 1.2,\dots$ (\fref{fig:FT_ex}c), and let us pick the unit cell convention such that both of the sites at $ \pm 0.2$ are assigned to the unit cell at $x=0$. 
Under $C_{2x}$, the orbitals are all invariant, i.e., $C_{2x}: \hat c_{i, x}^\dagger \mapsto \hat c_{i, x}^\dagger$ for $i = 1,2$. Under $C_{2y}$, however, the two orbitals are interchanged, giving $C_{2y}: \hat c_{1, x}^\dagger \leftrightarrow \hat c_{2,-x}^\dagger$. The momentum space representations are now given by
\begin{equation}\label{eq:}
\eqalign{
U_{k_x}(C_{2x}) =& \left(
\begin{array}{cc}
1 & 0 \\
0 & 1
\end{array}
\right),
U_{k_0}(C_{2y}) = \left(
\begin{array}{cc}
0 & 1 \\
1 & 0
\end{array}
\right)
}
\end{equation} 
for $k_0 = \Gamma, X$. Although $U_{k_0} (C_{2y})$ interchanges the two sites, it can be readily diagonalized, and we see that the two states furnish the representation $A\oplus B_{3}$ at both $\Gamma$ and $X$.

As our last example, let us replace the $C_{2x}$ rotation by the $2_{1}$ screw $S_{x}\equiv \{ C_{2x}| \frac{1}{2} \vechat x\}$, but keep the $C_{2y}$ symmetry (\fref{fig:FT_ex}d). Note that the point group is still $D_2$ even when a rotation is replaced by a screw.
To respect $S_{x}$, we need two orbitals in each unit cell, located respectively at the integer and the half-integer positions.
We consider $s$ orbitals such that the action of $C_{2y}$ is simply $\hat c_{1,x} \mapsto \hat c_{1,-x}$ for the integer sites, and $\hat c_{2,x}^\dagger \mapsto \hat c_{2,-x-1}^\dagger$ for the half-integer sites, as in equation \eref{eq:half_int_sites}. 
Repeating the preceding analysis, one finds
\begin{equation}\label{eq:}
U_{\Gamma}(C_{2y}) = \left(
\begin{array}{cc}
1 & 0 \\
0 & 1
\end{array}
\right);~~~
U_{X}(C_{2y}) = \left(
\begin{array}{cc}
1 & 0 \\
0 & -1
\end{array}
\right).
\end{equation} 
The screw symmetry, however, will interchange the two sets of sites:
\begin{equation}\label{eq:}
S_{x}: \hat c_{1,x} \mapsto \hat c_{2,x} ~\&~\hat c_{2,x} \mapsto \hat c_{1,x+1}.
\end{equation} 
Notice $S_x^2: \hat c_{i,x} \mapsto \hat c_{i,x+1} $ for $i=1,2$, consistent with the fact that $S_x^2 = \{ E| \vechat x\}$, the unit translation. 
Upon Fourier transforms, one finds
\begin{equation}\label{eq:}
S_x: \hat c_{1,k_x} \mapsto \hat c_{2,k_x} ~\&~\hat c_{2,k_x} \mapsto \hat c_{1,k_x} e^{-i k_x},
\end{equation} 
which corresponds to the matrix
\begin{equation}\label{eq:}
U_{k_x}(S_x) = \left(
\begin{array}{cc}
0 & e^{-ik_x} \\
1 & 0
\end{array}
\right).
\end{equation} 
If we focus at the $\Gamma$ point, we see that the eigenvalues of $U_{\Gamma}(S_x)$ is $\pm 1$, and we can identify the representation as $A\oplus B_2$ in terms of the irreps of $D_2$. This is possible, because at $\Gamma$ all translations $\{ E| \vec a\}$ are represented as $\id$, and so the translation part of $S_x$ does not matter. At $X$, however, the eigenvalues of $U_{X}(S_x)$ is $\pm i$, which is consistent with $U_{X}(S_x^2) = U_{X}(\{ E|\vechat x\}) = - \id$. Because of this, one can no longer identify the representations at $X$ using the linear representations of $D_2$. In addition, if the system has time-reversal symmetry, the two states at $X$ will become degenerate as their screw eigenvalues form a conjugate pair. In this case, the band crossing in \fref{fig:CR}c will be pinned at the $X$ point.

In all the examples before, we have left the discussion on $C_{2z}$ implicit. This is because the representation for $C_{2z}$ is known once those for $C_{2x}$ and $C_{2y}$ are given. Since we have replaced $C_{2x}$ by $S_x$ in the current example, the representation for $C_{2z}$ does not follow obviously and deserves a separate discussion. First, we note that there is no longer a point-group origin in the unit cell, and as a result the rotation centers for $C_{2y}$ and $C_{2z}$ no longer coincide. We have indicated the two inequivalent $C_{2z}$ centers in each unit cell by crosses in \fref{fig:FT_ex}d. Let us consider $C_{2z}$ at the rotation center immediately to the right of the origin, which is given explicitly by $\{ C_{2z} | \frac{1}{2} \vechat x\}$ and it leaves the point $x=\frac{1}{4}$ invariant. Note that despite the appearance of the fractional translation, this is a regular rotation symmetry because $\{ C_{2z} | \frac{1}{2} \vechat x\}^2 = \{ E|\vec 0\}$.
Its action on the fermions is given by: $C_{2z}: \hat c_{1,x} \leftrightarrow  \hat c_{2,-x} $.
Correspondingly, its momentum space representation is, for $k_0 = \Gamma, X$, 
\begin{equation}\label{eq:}
U_{k_0}(C_{2z}) = \left(
\begin{array}{cc}
0 & 1 \\
1 & 0
\end{array}
\right).
\end{equation}  
While one can check that $U_{X}(S_x)U_{X}(C_{2y})  = U_{X}(C_{2z})$, unlike the previous cases,  $U_{X}(C_{2y})$ and $U_{X}(S_x)$ do not commute, and one finds  $U_{X}(C_{2y}) U_{X}(S_x) =- U_{X}(C_{2z})$. This is a manifestation of the corresponding non-commutativity of the symmetry elements:
\begin{equation}\label{eq:}
\eqalign{
\{ C_{2x} | \vechat x/2\} \{ C_{2y} | \vec 0\}  =& \{ C_{2z} | \vechat x/2\};\\
\{ C_{2y} | \vec 0\} \{ C_{2x} | \vechat x/2\}  =& \{ C_{2z} | -\vechat x/2\}
= \{ E | -\vechat x\} \{ C_{2z} | \vechat x/2\}.
}
\end{equation}

\section{Space of band structures \label{sec:BS}}
Our discussion so far has closely followed the classic approaches for deriving and analyzing the symmetry representation of electronic states within band theory. Our next step is to integrate the modern ideas of band topology, especially insights from the K-theoretic approach to classifying topological (crystalline) insulators, to reformulate the problem of symmetries in band structures. The key idea is that, in terms of symmetry representations, the set of all possible band structures can be viewed as the abelian group $\mathbb Z^{d_{\rm BS}}$, which is very similar to a finite dimensional vector space.

\subsection{Compatibility relations as linear constraints
\label{sec:CR_lin}}
As we have seen, any representation of $\mathcal G_{\vec k}$ can be written as a direct sum of the irreps. For instance, we have encountered various two-dimensional representations of the point group $D_2$, like $A\oplus B_{3}$, $B_{3} \oplus B_{3}$, etc. If we pick a convention in which the irreps are arranged in, say, the ascending order in terms of the energies of the associated band, then $A\oplus B_3$ and $B_3 \oplus A$ correspond to different band orderings. In studying the topological properties of a set of bands, however, we are not interested in the energy ordering of the bands within the set. Instead, it suffices to simply count the number of times each irrep appears in the set of bands \cite{Ari, PhysRevX.7.041069}. For instance, we can denote any representation of $D_2$ by the four non-negative integers corresponding to the multiplicities of the irreps $(n^{A}, n^{B_1}, n^{B_2}, n^{B_3})$:
\begin{equation}\label{eq:}
\begin{array}{rl}
A \oplus B_3 &\mapsto (1,0,0,1);\\
B_3 \oplus B_3 &\mapsto  (0,0,0,2);\\
A \oplus B_2 \oplus A& \mapsto (2,0,1,0),
\end{array}
\end{equation} 
etc.

The compatibility relations become simple linear relations on the multiplicities of irreps. For concreteness, consider the relations listed in equation \eref{eq:D2_CR_spinless}, which encodes how the representations at $\Gamma$ subduces to those along the $\Sigma$ line $(k_x,0,0)$. A representation of $\mathcal G_{\Sigma}/T$ is specified by two integers $(n_{\Sigma}^{A}, n_{\Sigma}^{B})$, but these two integers are fully specified by the representation assigned at the higher-symmetry point $\Gamma$. As an example, if $U_\Gamma = B_1 \oplus B_2$, we know immediately that $U_\Sigma = B \oplus B$. In terms of the irrep multiplicities, we have the relations
\begin{equation}\label{eq:lin_CR}
\eqalign{
n_{\Sigma}^{A} &= n_{\Gamma}^{A} + n_{\Gamma}^{B_3};\\
n_{\Sigma}^{B} &= n_{\Gamma}^{B_1} + n_{\Gamma}^{B_2}.
}
\end{equation} 
The same relation holds if we replace $n_{\Gamma} $ on the right hand side by $n_{X}$, and so, one can conclude that a set of bands isolated from above and below along $\Sigma$ must have the representation multiplicities satisfying 
\begin{equation}\label{eq:lin_CR}
\eqalign{
n_{\Gamma}^{A} + n_{\Gamma}^{B_3} &= n_{X}^{A} + n_{X}^{B_3};\\
n_{\Gamma}^{B_1} + n_{\Gamma}^{B_2} &= n_{X}^{B_1} + n_{X}^{B_2}.
}
\end{equation} 
One can check that these equalities are satisfied by the representations in \fref{fig:CR}a, but violated in \fref{fig:CR}b, consistent with whether or not the bands are isolatable. 

Strictly speaking, all the irrep multiplicities are non-negative integers. We will relax this condition and let these multiplicities take any value in $\mathbb Z$. While the negative entries are not physical, they can be viewed as the formal inverses of the positive ones under the direct sum operation, and by including them we can view the irrep multiplicities as elements of an abelian group. This approach is inspired by the K-theoretic treatment of the classification of topological band insulators  \cite{PhysRevLett.95.016405, Kitaev, Freed2013, PhysRevB.95.235425, PhysRevX.7.041069, shiozaki2018atiyahhirzebruch}, and, as we will see, it greatly simplifies the analysis of the solutions to compatibility relations.

\subsection{Solution space for the compatibility relations}
Since the compatibility relations are simply linear constraints on the irrep multiplicities, we can handle them using linear algebra, i.e., to encode the compatibility relations by a matrix $\mathcal C$. For instance, we can express equation \eref{eq:lin_CR} as the system of linear equations
\begin{equation}\label{eq:}
\left(
\begin{array}{cccccc}
1 & 0 & 0 & 1 & -1 & 0 \\
0 & 1 & 1 & 0 & 0 & -1 
\end{array}
\right)
\left(
\begin{array}{c}
n_{\Gamma}^{A}\\
n_{\Gamma}^{B_1}\\
n_{\Gamma}^{B_2}\\
n_{\Gamma}^{B_3}\\
n_{\Sigma}^{A}\\
n_{\Sigma}^{B}
\end{array}
\right)
= \vec 0.
\end{equation} 
Schematically, we can write this linear equation as $\mathcal C \vec n = \vec 0$.
By the same token, we can also augment the linear space to include the multiplicities $n_{X}$'s, and at the same time enlarge the matrix $\mathcal C$ to include the compatibility relations relating $X$ and $\Sigma$.

Given a fixed symmetry setting, i.e., a space group, the absence or presence of time-reversal symmetry, and the specification of spinless vs. spinful electrons, one can collect all the irrep multiplicities at all the high-symmetry momenta into a vector $\vec n$, and the compatibility relations into a matrix $\mathcal C$\footnote{
Technically, $\vec n$ is not really a vector because it is integer-valued, and $\mathbb Z$ is not a field. We will nevertheless use the word ``vector'' loosely because the linear algebra we need is very similar to the ones we encounter for vector spaces.
}. Once this is achieved, we simply look for the solutions to the equation $\mathcal C \vec n = \vec 0$, which represents combinations of symmetry representations which globally satisfy all the symmetry constraints for an isolated set of bands. 

However, one must be careful in ensuring that all constraints are taken into account. As an example, let us consider a spinless two-dimensional system defined on the $x$-$y$ plane, and suppose that the spatial symmetries are described by the wallpaper group $p2$ (no.\ $2$), which is generated by the two lattice translations and the two-fold rotation $C_{2z}$. There are only four high-symmetry momenta: $\Gamma =(0,0)$, $X=(\pi,0)$, $Y=(0,\pi)$, and $M=(\pi,\pi)$. At each of the momenta, the little group is the entire space group. Since there is no nonsymmorphic elements, we can simply label the representations by that of the point group $C_2$. The irreps, as shown in \tref{tab:C2_spinless}, are either even ($A$) or odd ($B$) under $C_{2z}$. It then appears that all the symmetry data is encoded in the eight integers $( n_{\vec k_0}^{ A}, n_{\vec k_0}^{ B} )$ for $\vec k_0 = \Gamma, X,Y,M$. Since $C_{2z}$ is broken along any line joining any pair of these special momenta, it would appear that there are no compatibility relations, and that the space of possible band structures would be $\mathbb Z^8$, i.e., $8$ independent integers. But this is wrong, because we have neglected the condition that the number of bands in an isolated set, $\nu$, is also a topological quantity, and so we actually have the four compatibility relations
\begin{equation}\label{eq:}
n_{\vec k_0}^{ A}+ n_{\vec k_0}^{ B}  = \nu.
\end{equation} 
Taking this into account, the space of band structure is determined by 
\begin{equation}\label{eq:p2_n}
\eqalign{
\vec n &= (n_{\Gamma}^{A}, n_{\Gamma}^{ B}, n_{X}^{ A}, n_{X}^{ B}, n_{Y}^{ A}, n_{Y}^{ B}, n_{M}^{ A}, n_{M}^{ B}, \nu )^T;\\
\mathcal C &=
\left(
\begin{array}{ccccccccc}
1 & 1 & 0 & 0 & 0 & 0 &0 & 0 & -1\\
0 & 0 & 1 & 1 & 0 & 0 &0 & 0 & -1\\
0 & 0 & 0 & 0 & 1 & 1 & 0 & 0 & -1\\
0 & 0 & 0 & 0 &0 & 0 & 1 & 1 & -1
\end{array}
\right).
}
\end{equation} 
Here, $\vec n \in \mathbb Z^9$, and to obtain a band structure we subject it to the  four linearly independent compatibility relations. This implies the solutions to $\mathcal C \vec n = \vec 0$ can be characterized by $5$ integers. In other words, it can be identified with $\mathbb Z^5$ \cite{PhysRevX.7.041069, Po2017}.
Denoting the solution space by $\{ {\rm BS} \}$, we can write it as
\begin{equation}\label{eq:}
\{ {\rm BS} \} = \left\{ \sum_{i=1}^5m_i \vec b_i ~:~ m_i \in \mathbb Z \right\},
\end{equation} 
where we can choose the $5$ basis as
\begin{equation}\label{eq:b_vec}
\eqalign{
\vec b_1 &= (1,0,1,0,1,0,1,0,1)^T;\\
\vec b_2 &= (0,1,0,1,0,1,0,1,1)^T;\\
\vec b_3 &= (1,0,0,1,1,0,0,1,1)^T;\\
\vec b_4 &= (1,0,1,0,0,1,0,1,1)^T;\\
\vec b_5 &= (0,0,0,0,0,0,1,-1,0)^T.
}
\end{equation} 
This concludes our analysis for the wallpaper group $p2$ assuming spinless electrons. Note that this is among the simplest symmetry settings. Readers who would want to see a similar analysis with a slightly more complicated setup are encouraged to consult \cite{PhysRevX.7.041069}, which provided a detailed analysis for the wallpaper group $p4mm$ (no.\ 11).

Having worked through a concrete example, one could imagine performing the same analysis for any symmetry setting. The general procedures are: 
\begin{enumerate}
\item Compile the list of all types of high-symmetry momenta $\vec k$, and for each of them list out all the possible irreps, indexed by $\alpha$, which are appropriate for the symmetry setting (spinless vs.\ spinful electrons). We can then aggregate all these irrep multiplicities $n_{\vec k}^{\alpha}$, together with the total number of bands $\nu$, into an integer-valued vector $\vec n$. We let $D$ denote the number of integers involved, i.e., $\vec n \in \mathbb Z^D$.
\item Find all compatibility relations between the entries of $\vec n$. Let $\tilde D$ denote the number of such relations, then assemble them into the $\tilde D \times D$ matrix $\mathcal C$.
\item Compute the right null vectors of the matrix $\mathcal C$, which spans the solution space $\{ {\rm BS}\} \simeq \mathbb Z^{d_{\rm BS}}$\footnote{A quick proof that the solution space can be identified with $\mathbb Z^{d_{\rm BS}}$ with $d_{\rm BS}\leq D$: we may view $\mathcal C$ as a map $\mathcal C: \mathbb Z^{D} \rightarrow \mathbb Z^{\tilde D}$. The solution space is simply $\ker \mathcal C$, which is a subgroup of $\mathbb Z^{D}$. Any subgroup of $\mathbb Z^{D}$ is isomorphic to $\mathbb Z^{d}$ for some $d\leq D$.
}.
\end{enumerate}
In the example above, we simply have $d_{\rm BS} = D - \tilde D$. This is because all the compatibility relations we listed are linearly independent. In the general case, however, one may need to analyze the matrix $\mathcal C$ more carefully to identify the number of independent null vectors. 

Although the construction of $\mathcal C$ is straightforward for the simple example we studied, for a more general setting one has to also taken into account the constraints on representations at symmetry-related momenta, as well as the band connectivity in the presence of nonsymmorphic symmetries. Furthermore, we have ignored the action of time-reversal symmetry in this section. As we discussed, time-reversal could constrain the representations in three different ways: when it does not change the irrep, no additional compatibility relation is generated; when it pairs two distinct irreps, say $u^\alpha_{\vec k_0}$ and $u^\beta_{\vec k_0}$, we should add a relation $n_{\vec k_0}^\alpha = n_{\vec k_0}^\beta$; when it pairs an irrep with another copy of itself, however, we can no longer treat it as a linear constraint. Specifically, if the irrep $\alpha$ of $\mathcal G_{\vec k_0}$ is paired with itself under time-reversal, the multiplicity $n_{\vec k_0}^\alpha$ must be an even integer in any band structure. While this is not a linear constraint, we can handle it in our formalism by rewriting $n_{\vec k_0}^\alpha  = 2 \tilde n_{\vec k_0}^\alpha$, and then re-express all the compatibility relations in terms of $\tilde n_{\vec k_0}^\alpha \in \mathbb Z$. With this modification, the discussed linear structure is maintained even for systems with time-reversal symmetry.

In principle, to correctly identify the solution space $\{ {\rm BS}\}$ it is crucial that no compatibility relation is missed. Otherwise, one might end up with a representation vector $\vec n$ which satisfies all the listed relations, but actually violated a missed one and is therefore not isolatable. The exhaustive computation for all compatibility relations can become quite involved for complicated three-dimensional space groups with a rich set of symmetries and a non-primitive Bravais lattice. Curiously, through the theory of symmetry indicators, we can prove that the exhaustive list of all symmetry constraints in a band structure can be exposed by systematically analyzing the space of atomic insulators.

\section{Space of atomic insulators \label{sec:AI}}
Although the notion of compatibility relations, which is tied to the symmetry representations of the Bloch states, is inherently a concept defined in the momentum space, a special class of solutions for the relations can be constructed in the real space. These are the atomic insulators, which arise when a set of symmetry-related real-space orbitals are completely filled. 
Being insulators, they automatically give rise to a band structure, defined as bands that can be isolated from all others by a continuous band gaps at all high symmetry momenta.
As these states are, by definition, smoothly connected to an entanglement-free ground state, they only correspond to a special class of band structures. Nevertheless, they provide important information on the general structure of the solution space $\{{\rm BS}\}$, as we will see below.

\subsection{Atomic insulators in the real space}
An atomic insulator arises when a set of real-space orbitals are completely filled. In terms of their symmetry properties, they are specified by two pieces of data: (i) the lattice, and (ii) the orbital characters on each site. 
As the lattice respects the spatial symmetries, if a site is located at a point $\vec x$ in space, then there must also be a site at all the points $\{ g(\vec x) ~:~ g \in \mathcal G\}$. This is called the crystallographic orbit of $\vec x$.
Some symmetries may leave the point $\vec x$ invariant, and we define the site symmetry group $\mathcal G_{\vec x} \equiv \{ g \in \mathcal G~:~ g(\vec x) = \vec x\}$. This is conceptually similar to the notion of little group in the momentum space, but with the crucial difference that a lattice translation is never going to leave any site invariant, and so the site symmetry group never contains any lattice translations. As such, $\mathcal G_{\vec x}$ is always a subgroup of the point group.
From the symmetry action of $\mathcal G$, we can partition the real space into Wyckoff positions, which can be viewed as a classification of the distinct types of 
crystallographic orbits. Any point in space belongs to a Wyckoff position, which is labeled by the alphabets a,b,c,\dots. As such, in terms of symmetries, we can describe a lattice by specifying the Wyckoff positions the sites belong to. Note that a Wyckoff position may contain free parameters, and in that case the sites in the Wyckoff positions could move around in space while respecting all symmetries (corresponding to the tuning of the free parameters). When these parameters are tuned to special values, the sites may collapse at a higher-symmetry point in space, which leads to a lattice realizing a higher-symmetry Wyckoff position.

\begin{figure}[h]
\begin{center}
{\includegraphics[width=0.68 \textwidth]{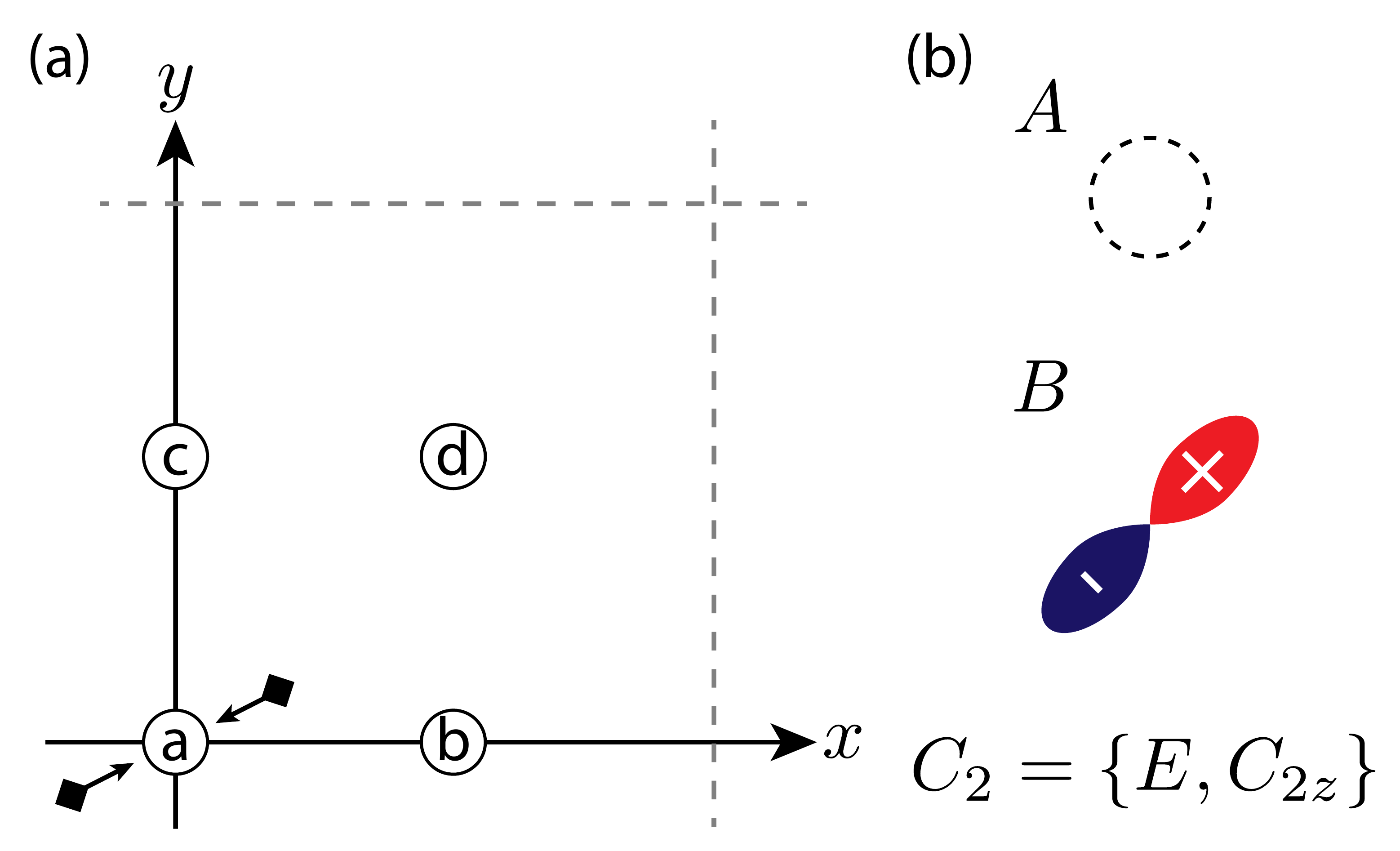}} 
\caption{Examples of Wyckoff positions and orbitals. (a) A two-dimensional system with $C_2$ site symmetry at the four Wyckoff positions labeled by a through d. The dashed lines indicate the unit cell. The sites in Wyckoff position e are represented by filled diamonds. As shown by the arrows, these sites can be smoothly moved towards the site in Wyckoff position a without breaking any symmetry. (b) The orbital types for the Wyckoff positions a through d can be labeled using the irreps of the group $C_2$.
\label{fig:Lattice}
 }
\end{center}
\end{figure}

As a concrete example, consider again a two-dimensional system with spatial symmetries described by the wallpaper group $p2$ (no.\ 2). There are four highest-symmetry points in each unit cell, which correspond to the four Wyckoff positions (\fref{fig:Lattice}a):
\begin{equation}\label{eq:}
\eqalign{
&\mathcal W_{\rm a}: (0,0);~~
\mathcal W_{\rm b}: (1/2,0);\\
&\mathcal W_{\rm c}: (0,1/2);~~
\mathcal W_{\rm d}: (1/2,1/2).
}
\end{equation} 
For each of these four Wyckoff positions, the site symmetry group is the full point group $C_2= \{ E,C_{2z}\}$.
In contrast, $\mathcal W_{\rm e} $ with representative sites $(x,y)$ and $(-x,-y)$ is the general position, and it has a trivial site symmetry group. Indeed, by setting the free parameter to, say, $x, y= 0$, we can collapse the two sites in each unit cell to obtain  $\mathcal W_{\rm a}$, or similarly to any of the other high-symmetry Wyckoff positions.

After specifying the lattice sites, we have to specify the orbitals which are filled. In the absence of spin-orbit coupling, atomic orbitals can be labeled by their azimuthal quantum numbers $\ell = s,p,d,\dots$. In a crystallographic environment, however, the orbital degeneracy is generally lifted due to crystal field splitting. Instead, the appropriate symmetry labels for orbitals centered at a site $\vec x$ will be the irreps of the site symmetry group $\mathcal G_{\vec x}$. 
For instance, on each site in the Wyckoff positions a through d shown in \fref{fig:Lattice}a, we can label the orbitals by the two irreps of $C_2$ in \tref{tab:C2_spinless} when the system is spinless. Representative orbital shapes are shown in \fref{fig:Lattice}b. Note that an example of the $B$ irrep, which is odd under $C_{2z}$, can be obtained by any linear superposition of the $p_x$ and $p_y$ orbitals. In contrast, a $p_z$ orbital transforms trivially under $C_{2z}$, and is indistinguishable from an $s$ wave orbitals with the symmetries at hand.

Given the full list of Wyckoff positions and irreps of the site-symmetry group, which are both well-documented in the literature \cite{ITC, Bradley, Bilbao_rep}, one can exhaustively construct all possible atomic insulators in any given symmetry setting. In particular, as the composite of two atomic insulators is just another atomic insulator (with the total electron filling being the sum of the two), it suffices to study the ``elementary ones'' which serve as the building blocks for all other atomic insulators. 
For instance, from the discussion above we can readily infer that any atomic insulator in the wallpaper group $p_2$ (no. 2) can be viewed as a stack of the eight atomic insulators obtained by combing one of the four Wyckoff positions $\mathcal W_{\rm a,b,c,d}$ with one of the two $C_2$ irreps $A$ or $B$. In principle, one could also consider the atomic insulator obtained by filling an orbital localized to a site in the general position $\mathcal W_{\rm e}$. However, given we could choose the free parameter to collapse the sites to the higher-symmetry Wyckoff positions, any atomic insulator constructed from $\mathcal W_{\rm e}$ would automatically be obtainable from stacks of ones defined on, say, $\mathcal W_{\rm a}$, and so we do not have to worry about such Wyckoff positions.

We have given an example in which we can construct any possible atomic insulator based on a certain collection of elementary ones. This is the underlying principle behind the theory of band representation developed in Refs.\ \cite{PhysRevLett.45.1025, PhysRevB.23.2824, PhysRevLett.61.1005, Bacry, MichelZak}, which provides a framework for identifying such building blocks and analyzing the topological equivalence between atomic insulators.
While we are also interested in the space of all possible atomic insulators, for our purpose it is unimportant to identify which ones are elementary; instead it is more economical to simply work with a possibly redundant set, and then extract the linearly independent data using linear algebra.

\subsection{Representation vectors of atomic insulators}
After specifying an atomic insulator in the real space, we can obtain its symmetry representations in the momentum space following the Fourier transform discussed in \sref{sec:FT}.
So, for every specification of lattice and filled orbitals in the real space, we obtain a corresponding representation vector $\vec a$. Having specified a symmetry setting, there will generally be some $d_{\rm AI}$ basis vectors, and we identify the space of atomic insulators as
\begin{equation}\label{eq:}
\{ {\rm AI} \} \equiv \left \{ \sum_{i=1}^{d_{\rm AI}} m_i \vec a_i~:~ m_i \in \mathbb Z \right\},
\end{equation} 
which is abstractly the group $\mathbb Z^{d_{\rm AI}}$. Importantly, the atomic insulators are simply special solutions to all the compatibility relations, and so we can conclude 
\begin{equation}\label{eq:}
\{ {\rm AI} \} \leq \{ {\rm BS}\},
\end{equation} 
i.e., $\{ {\rm AI} \} $ is a subgroup of $ \{ {\rm BS}\}$.

Let us compute the basis for $\{ {\rm AI}\}$ explicitly for the $p2$ example.
In the notation of equation \eref{eq:p2_n}, the representation vectors of the eight atomic insulators mentioned are
\begin{equation}\label{eq:p2_AI_vecs}
\eqalign{
\vec a_{{\rm a}}^{A} &= (1,0,1,0,1,0,1,0,1)^T;\\
\vec a_{{\rm a}}^{B} &= (0,1,0,1,0,1,0,1,1)^T;\\
\vec a_{{\rm b}}^{A} &= (1,0,0,1,1,0,0,1,1)^T;\\
\vec a_{{\rm b}}^{B} &= (0,1,1,0,0,1,1,0,1)^T;\\
\vec a_{{\rm c}}^{A} &= (1,0,1,0,0,1,0,1,1)^T;\\
\vec a_{{\rm c}}^{B} &= (0,1,0,1,1,0,1,0,1)^T;\\
\vec a_{{\rm d}}^{A} &= (1,0,0,1,0,1,1,0,1)^T;\\
\vec a_{{\rm d}}^{B} &= (0,1,1,0,1,0,0,1,1)^T.\\
}
\end{equation} 
One can verify that each of these vectors satisfies $\mathcal C \vec a_{w}^\alpha = \vec 0$ for the compatibility relation matrix in equation \eref{eq:p2_n}.
These eight vectors, however, are not linearly independent. For instance, one can check that\footnote{
There is a somewhat deeper reason for this: As we have seen, an atomic insulator built using the general position of $p_2$ can be smoothly deformed into a stack of atomic insulators with electrons localized to any one of the four high-symmetry Wyckoff positions. Since the two sites in the general positions are related by the $C_{2z}$ symmetry, the even and odd combinations of the orbitals on these two sites respectively correspond to the $A$ and $B$ irreps of $C_2$. This establishes an equivalence between the atomic insulators, which is then reflected in the representation vectors.
}
\begin{equation}\label{eq:}
\vec a_{w}^{A}+\vec a_{w}^{B}
= (1,1,1,1,1,1,1,1,2)^T,
\end{equation} 
for $w = {\rm a,b,c,d }$. 
Using this relation, we can choose to eliminate $\vec a_{{\rm b}}^{B}$, $\vec a_{{\rm c}}^{B}$, and $\vec a_{{\rm d}}^{B}$ from the basis set, and we see that $d_{\rm AI} = 5$. This gives a possible choice of the basis vectors
\begin{equation}\label{eq:}
\eqalign{
\vec a_{1} &= \vec a_{{\rm a}}^{A};~~
\vec a_{2} = \vec a_{{\rm a}}^{B};~~
\vec a_{3} = \vec a_{{\rm b}}^{A};~~
\vec a_{4} = \vec a_{{\rm c}}^{A};\\
\vec a_{5} &= \vec a_{{\rm a}}^{A}-\vec a_{{\rm b}}^{A}-\vec a_{{\rm c}}^{A}+\vec a_{{\rm d}}^{A}.
}
\end{equation} 
One may wonder why we choose such a specific combination in $\vec a_5$, as it would have been more natural to simply use $\vec a_{{\rm d}}^{A}$. We first note that this is a legitimate choice, since 
\begin{equation}\label{eq:}
\eqalign{
&\left(
\begin{array}{ccccc}
\vec a_1 & \vec a_2 & \vec a_3 & \vec a_4 & \vec a_5
\end{array}
\right)\\
&=
\left(
\begin{array}{ccccc}
\vec a_{{\rm a}}^{A} & \vec a_{{\rm a}}^{B}& \vec a_{{\rm b}}^{A} & \vec a_{{\rm c}}^{A} & \vec a_{{\rm d}}^{A}\end{array}
\right)
\left(
\begin{array}{ccccc}
1 & 0 & 0 & 0 & 1\\
0 & 1 & 0 & 0 & 0\\
0 & 0 & 1 & 0 & -1\\
0 & 0 & 0 & 1 & -1\\
0 & 0 & 0 & 0 & 1
\end{array}
\right).
}
\end{equation} 
The matrix here is unimodular, i.e., invertible over the integers, and so the two sets of basis vectors are related by a basis transformation. 
The rationale behind our choice will become apparent once we write it out explicitly in terms of the irrep multiplicities:
\begin{equation}\label{eq:p2_a5}
\vec a_5 = (0,0,0,0,0,0,2,-2,0)^T,
\end{equation} 
and by comparing with equation \eref{eq:b_vec} we can conclude $\vec a_i = \vec b_i$ for $i=1,\dots, 4$ and $\vec a_5 = 2 \vec b_5$. This is a precursor for the definition and computation for the symmetry indicator group.

Let us close this section by describing the general procedures for computing the group $\{ {\rm AI}\}$ for any symmetry setting:
\begin{enumerate}
\item Identify all the Wyckoff positions which cannot be promoted to a higher-symmetry one by choosing specific values of the free parameters, if any.
\item For each of the Wyckoff positions of interest, consider all the possible irreps of the site symmetry group.
\item Construct the representation vector for each combination of Wyckoff position and site symmetry group irrep through Fourier transform, and then extract the $d_{\rm AI}$ linearly independent basis vectors. This gives $\{ {\rm AI} \} \simeq \mathbb Z^{d_{\rm AI} }$.
\end{enumerate}

\subsection{Fragile topology \label{sec:fragile}}
By definition, the group $\{ {\rm AI}\}$ is obtained by allowing all possible addition and subtraction between the representation vectors arising from atomic insulators. In order to interpret an element $\vec a \in \{ {\rm AI}\}$ as the irrep multiplicities of any physical band structure, we need to impose an additional condition that all the entries are non-negative. This is called the ``physical condition'' in \cite{Po2017}. Let us now pose the question: are all physical entries of $\{ {\rm AI}\}$ realizable by some atomic insulator?

Clearly, the vectors obtained by Fourier transforming an atomic insulator defined in the real space, like those listed in equation \eref{eq:p2_AI_vecs}, would be realizable. When two such atomic insulators are stacked, i.e., there are two inequivalent sets of filled orbitals in the real space, the representation vector of the composite will be simply the sum of the two individual ones. Since the physical stacking only leads to addition, but not subtraction, there could be elements of $\{ {\rm AI}\}$ which are never realizable by physical atomic insulators \cite{Po2017}. In fact, some of the filling-enforced quantum band insulators discussed in \cite{Poe1501782} are examples of such possibilities; there, the non-atomic filling is realized precisely by an effective subtraction between the electron fillings of certain atomic insulators. When a physical element (i.e., entries are non-negative) of $\{ {\rm AI}\}$ is not realizable by a physical atomic insulator, it also signals a topologically nontrivial state. In particular, it could correspond to a peculiar form of band topology,  dubbed ``fragile topology'' in \cite{PhysRevLett.121.126402}. The defining feature of fragile topological bands is that, if they are the only filled bands, there will a topological obstruction for symmetrically deforming the ground state into a product state. However, the obstruction can be resolved upon the addition of some auxiliary filled bands which are in a suitable atomic limit. Intuitively, such fragile states arise from the effective subtraction between atomic insulators, which leads to a physical state that by itself is non-atomic. Once the subtrahend, an atomic state, is added to the fragile state, the topological obstruction is resolved.

Although we have motivated the existence of fragile topology from the physicality of elements in $\{{\rm AI}\}$, it is a more general concept and some fragile bands may not be diagnosable from symmetry representations \cite{PhysRevLett.121.126402, PhysRevLett.120.266401, bouhon2018wilson, PhysRevB.99.045140, PhysRevX.9.031003, song2019fragile,hwang2019fragile,song2019real,2020arXiv200203836E}.
The described defining feature of fragile topology should be contrasted with the band topology described by conventional topological invariants, like the Chern number, which are stable against the addition of atomic degrees of freedom. The peculiar nature of fragile topology suggests that it cannot be analyzed within a conventional K-theory framework, and it was pointed out that such phases are better viewed as elements of a commutative monoid \cite{song2019fragile}.  
It is also an open question whether or not fragile topological states can have robust physical properties, like surface states, which are not possible in any atomic insulator.
We also mention that the notion of a fragile topological insulator can also be defined in an interacting system \cite{PhysRevB.99.125122}, and is even realized in graphene-based heterostructures \cite{PhysRevX.8.031089, PhysRevLett.123.036401, PhysRevB.99.195455, PhysRevX.9.021013, zaletel2019gatetunable}.

\section{Symmetry indicators \label{sec:SI}}
Having constructed $\{ {\rm BS}\}$ and $\{ {\rm AI}\}$, corresponding respectively to the general and special solutions to the compatibility relations for obtaining an isolatable set of bands, we are now in the position to compare them, which is equivalent to asking if the inclusion $\{ {\rm AI}\} \leq \{{\rm BS}\}$ is strict. 
As we have alluded to in the overview section, this comparison is mathematically achieved by computing the quotient group 
\begin{equation}\label{eq:}
X_{\rm BS} \equiv \frac{\{ {\rm BS}\}}{\{ {\rm AI}\}}.
\end{equation} 
In the following, we will attempt to unpack this mathematical definition, and provide a concrete recipe for computing the quotient group in our context.

\subsection{Smith normal form and the symmetry indicator group
\label{sec:SNF}}
As we have emphasized multiple times, the theory of symmetry indicators exploits the linear structure behind the symmetry constraints to enable the efficient diagnosis of topological band structures. To appreciate the importance of linearity, observe that, for any $\vec a_i \in \{ {\rm AI} \}$ and $q \in \mathbb Q$, the rational numbers, we automatically have 
\begin{equation}\label{eq:}
\mathcal C(q \vec a_i) = q (\mathcal C \vec a_i) = \vec 0.
\end{equation} 
For a general rational number $q$, the entries in $q \vec a_i$ may not be integer-valued, i.e., $q \vec a_i \not \in \mathbb Z^D$. In this case, there is no clear physical interpretation for the vector $q \vec a_i$. However, if one can find a rational number $q_{i}$ such that $q_{i} \vec a_i \in \mathbb Z^D$, it will be an element of $\{ {\rm BS}\}$. This leads us to consider elements of $\{ {\rm BS}\}$ taking the form
\begin{equation}\label{eq:}
\vec n = \sum_{i=1}^{d_{\rm AI}} q_i  \vec a_i, ~~~ q_i \in \mathbb Q~~ \& ~~\vec n \in \mathbb Z^{D}.
\end{equation} 
Importantly, as $q_i$ is a rational number, $q_i \vec a_i$ does not have to be an element of $\{ {\rm AI}\}$. In other words, we can obtain representation vectors for topological band structures simply by multiplying elements of $\{ {\rm AI}\}$ by suitable rational coefficients. 

We have already seen one such example: the basis vector $ \vec a_5$ in equation \eref{eq:p2_a5} remains integer-valued when divide by $2$, and this gives a nontrivial element of $\{ {\rm BS}\}$ which is not in $\{ {\rm AI}\}$.
Once such a nontrivial element is found, we can generate infinitely many examples by considering, for instance, $\frac{1}{2} \vec a_5 + m \vec a_1$ for $m\in \mathbb Z$. These examples, however, differ only by the stacking of some trivial atomic insulators, and their mutual distinction is not of interest to the diagnosis of topological materials. As such, we will identify all these nontrivial elements as being in the same class. Further noting $2 \times (\frac{1}{2} \vec a_5 )= \vec a_5$ is trivial, and that $\vec b_i = \vec a_i$ for $i=1,\dots, 4$, we can conclude
\begin{equation}\label{eq:}
X_{\rm BS} = \mathbb Z_2
\end{equation} 
for the wallpaper group $p_2$ with spinless electrons.

The analysis above can be done generally by introducing the Smith normal form. In our context, we can think of the Smith normal form as an integer-valued version of the singular value decomposition, where the unitary matrices are replaced by unimodular matrices, and the ``singular values'' are all positive integers.
For concreteness, let us suppose we have, following the discussion in \sref{sec:AI}, obtained a set of $d$ representation vectors spanning $\{ {\rm AI}\}$. These $d$ vectors may not be linearly independent, but we always have $d\geq d_{\rm AI}$. Let us denote these column vectors by $\vec a'_{i}$.
We can aggregate these vectors into a $D\times d$ matrix
\begin{equation}\label{eq:}
\mathcal A' \equiv 
\left(
\begin{array}{cccc}
\vec a'_1 & \vec a'_2 & \dots & \vec a'_d
\end{array}
\right),
\end{equation} 
where we also have $D\geq d_{\rm AI}$.
While we will not prove it, one can find unimodular matrices $U$ and $V$ such that
\begin{equation}\label{eq:SNF}
\mathcal A' = U
\left(
\begin{array}{cc}
\Sigma & 0 \\
0 & 0
\end{array}
\right) V^{-1} ; ~~ 
\Sigma = {\rm diag}(s_1,s_2,\dots, s_{d_{\rm AI}}),
\end{equation} 
where $s_i$ are all positive integers. The diagonal matrix above is called the Smith normal form of $\mathcal A'$.
The matrix above is written in a block-matrix form, such that the $0$'s all have the dimensionality required. In particular, $U$ and $V$ are respectively $D$ and $d$ dimensional.
The unimodular matrix $V$ provides a basis for $\{ {\rm AI} \}$:
\begin{equation}\label{eq:}
\mathcal A' V = \left(
\begin{array}{ccccccc}
\vec a_1 & \vec a_2 & \dots & \vec a_{d_{\rm AI}} & 0 & \dots & 0
\end{array}
\right).
\end{equation} 
The vectors $\vec a_i$ here are simply the basis vectors for $\{ {\rm AI}\}$.

To extract the non-zero part of $\mathcal A'V$, define  $\mathcal A \equiv \left(
\begin{array}{cccc}
\vec a_1 & \vec a_2 & \dots & \vec a_{d_{\rm AI}}
\end{array}
\right)$.
Similarly, we should focus on the first $d_{\rm AI}$ columns of $U$, which gives the $D\times d_{\rm AI}$ dimensional submatrix
\begin{equation}\label{eq:}
\mathcal B = \left(
\begin{array}{cccc}
\vec b_1 & \vec b_2 & \dots & \vec b_{d_{\rm AI}}  
\end{array}
\right).
\end{equation} 
Note that $\mathcal B$ is also an integer-valued matrix.
With these definitions, equation \eref{eq:SNF} reduces to the matrix equation $\mathcal A = \mathcal B \Sigma$. Column-by-column, we have
\begin{equation}\label{eq:}
\vec a_i = \vec b_i s_i ~~~{\rm for}~~~ i = 1,\dots, d_{\rm AI}.
\end{equation} 
Hence, we see that whenever $s_i > 1$, $\vec b_i =  \vec a_i/ s_i$ is a nontrivial entry of $\{ {\rm BS} \}$, i.e., it corresponds to a band structure which cannot be obtained from any atomic limit. Furthermore, the vectors $\vec b_i$  provide $d_{\rm AI}$ basis vectors for $\{ {\rm BS}\}$.

We have seen that, starting from the (possibly over-complete) set of atomic insulator representation vectors, the computation of the Smith normal form immediately leads to the identification of a set of linearly independent basis vectors for $\{ {\rm AI}\}$ and $\{ {\rm BS}\}$, and when any $s_i>1$, we see that there are topological band structures which are diagnosable its symmetry representations. The only remaining question is whether or not the basis $\{ b_{i}~:~ i = 1,\dots, d_{\rm AI}\}$ is complete for $\{ {\rm BS}\}$. In other words, we have to answer if $d_{\rm BS}$, the dimension of the solution space for the compatibility relations, is larger than $d_{\rm AI}$. This problem was studied in \cite{Po2017}, and the conclusion is
\begin{equation}\label{eq:}
d_{\rm BS} = d_{\rm AI}
\end{equation} 
for electrons in all 230 space groups with or without spin-orbit coupling and/ or time-reversal symmetry. This was basically proven by an explicit calculation for all symmetry settings. With hindsight, it may be possible to find a simple proof using the recently developed frameworks for the classification of topological crystalline phases \cite{PhysRevX.7.011020, PhysRevX.8.011040, Xiong_2018, PhysRevB.96.205106, PhysRevB.97.115153, PhysRevB.99.115116, shiozaki2018generalized}. We will not pursue this task here, but rather accept it as an established fact. Given this equality, we can conclude the symmetry indicator group is always finite, and can be written as
\begin{equation}\label{eq:}
X_{\rm BS} = \mathbb Z_{s_1}\times \mathbb Z_{s_2}\times\dots \times \mathbb Z_{s_{d_{\rm BS}}}.
\end{equation} 
In particular, it is nontrivial if and only if any of the factors $s_{i}>1$ in equation \eref{eq:SNF}.
A nontrivial consequence of the finiteness of $X_{\rm BS}$ is that knowledge on the atomic insulators, defined in the real space, provides an exhaustive check on the possibility of isolating a set of bands in the momentum space, as we will see next.

\subsection{Diagnosing topological materials using symmetry indicators
\label{sec:TM}}
Given a symmetry setting (space group, relevance of spin-orbit coupling, and the presence or absence of time-reversal symmetry), we have seen how to construct the basis vectors of $\{ {\rm AI}\}$ and $\{ {\rm BS}\}$ and compute the symmetry indicator group $X_{\rm BS}$. Now we discuss how to use these mathematical objects to diagnose the topological nature of materials. We note that the method discussed below is essentially the one described in \cite{Tang2019_NP}, and very similar schemes have been proposed in \cite{Zhang2019, Vergniory2019}.

Our starting point will be a collection of high-symmetry momenta, labeled by $\vec k$, and the representation $U_{\vec k}$ of $\mathcal G_{\vec k}$ which leaves the Bloch Hamiltonian $H_{\vec k}$ invariant.
Let $\nu$ denote the average number of electrons per unit cell, such that we could focus our attention to the $\nu$ low-lying eigenstates of the Bloch Hamiltonian $\{ | \psi_{i, \vec k} \rangle ~:~ i = 1,\dots, \nu\}$. We can extract the symmetry representation furnished by these states of interest as in equation \eref{eq:u_rep}, and then express them in terms of the irrep multiplicities following the discussion in \sref{sec:CR_lin}. 
These symmetry data can then be aggregated in to a single vector $\vec n$.
Our goal is to answer two questions: (i) Is $\vec n\in \{ {\rm BS}\}$? (ii) If yes, is $\vec n\in \{ {\rm AI}\}$?
If the answer to (i) is negative, by definition the vector $\vec n$ must have violated some compatibility relations, and so it cannot correspond to an isolated set of bands. This implies the materials is a (semi-)metal with symmetry and filling-enforced gaplessness at some high-symmetry momenta.
We note that this check subsumes the detection of topological (semi-)metals based on the electron filling constraints established in \cite{PhysRevLett.117.096404}.
Alternatively, if $\vec n\in \{ {\rm BS}\}$, we further ask if its representation content is consistent with an atomic insulator. When the answer to (ii) is negative, we know that $\vec n$ cannot admit any atomic description even though its representation is compatible with a full gap at all high-symmetry momenta. This implies it is a topological material candidate.

Computationally, these questions can be answered by introducing the pseudo-inverses to the matrices $\mathcal A$ and $\mathcal B$ defined in \sref{sec:SNF}. Let  $\mathcal A^{-1}$ and  $\mathcal B^{-1}$ be $d_{\rm BS}\times D$ matrices satisfying
\begin{equation}\label{eq:}
\mathcal A^{-1} \mathcal A = \id;~~~\mathcal B^{-1} \mathcal B = \id.
\end{equation} 
The membership questions posed above can then be rephrased as
\begin{enumerate}
\item Is $\mathcal B^{-1} \vec n$ integer-valued?
\item If yes, is $\mathcal A^{-1} \vec n$ integer-valued?
\end{enumerate}
Given $\mathcal A = \mathcal B \Sigma$ and $\Sigma$ is invertible by definition, we immediately have $\mathcal A^{-1} = \Sigma^{-1} \mathcal B^{-1}$, which satisfies $\mathcal A^{-1} \mathcal A = \id$. Therefore, to answer these questions one simply has to perform a single matrix multiplication. In particular, when $\mathcal B^{-1} \vec n = (m_1, m_2, \dots, m_{d_{\rm BS}})$ is integer-valued, we can identify its corresponding class in $X_{\rm BS}$ by evaluating
\begin{equation}\label{eq:}
r_i \equiv m_i ~{\rm mod} ~ s_i, 
\end{equation} 
and we say the symmetry indicator is trivial if $r_i =0$ for all $i$.

The analysis above requires only the computation of symmetry representations at all the high-symmetry momenta, and then perform a single matrix multiplication. This should be contrasted with a more conventional approach in which the topological invariants are evaluated by computing suitable wave-function integrals. 
However, one should note that this symmetry-based diagnosis does not capture the detailed energetics of the material, i.e., although the representations for $\vec n$ are compatible with a continuous gap at all high-symmetry momenta whenever $\vec n \in \{ {\rm BS}\}$, there can nevertheless be accidental crossings or strong energy dispersion which renders  the material (semi-)metallic. Besides, simply knowing the symmetry indicators does not immediately convey information on the concrete form of the nontrivial topology displayed by the material. To answer this, one has to perform a systematic study relating the symmetry indicators to the topological phases of matter, which we will discuss in the next section.

\section{Examples of topological phases diagnosed from symmetry indicators}
In this section, we provide a survey for the relations between the symmetry indicators and topological phases.  We will begin with two specific examples, and then move on to discuss the general case of materials with time-reversal symmetry.

\subsection{Rotation symmetry indicator for the Chern number}
We have computed the symmetry indicator for the wallpaper group $p_2$ in \sref{sec:SNF}, and found it to be $\mathbb Z_2$. We will now show that this is equivalent to the invariant
\begin{equation}\label{eq:}
\xi =  \sum_{\vec k_0=\Gamma, X,Y,M} n_{\vec k_0}^B ~{\rm mod}~2,
\end{equation} 
which was introduced in \cite{PhysRevB.83.245132,Ari,PhysRevB.86.115112}, and is known to be equivalent to the parity of the Chern number.
To establish this claim, we first notice that $\xi = 0$ for all the atomic insulator representation vectors in equation \eref{eq:p2_AI_vecs}. Now, we note that the nontrivial element is generated by $ \vec b_5$,  given by $n_{M}^A = 1, n_{M}^B = -1$, and $0$ for all other entries. In other words, $\xi[\vec b_5] = 1$. Since any representation vector $\vec n$ with a nontrivial symmetry indicator can be written as $\vec n =\vec a + \vec b_5$ for some $\vec a \in \{ {\rm AI}\}$, and $\xi$ is additive, we can then conclude the symmetry indicator is simply $\xi$.

As was shown in  \cite{PhysRevB.86.115112}, in the presence of a $C_n$ rotation symmetry one can construct similar formulas which diagnose the Chern number modulo $n$. Indeed, for the wallpaper group $p3$, $p4$ and $p6$, one finds the symmetry indicator groups $\mathbb Z_3$, $\mathbb Z_4$,  and $\mathbb Z_6$ respectively when we assume spinless electrons without time-reversal symmetry.

One can, however, imagine performing the same analysis for spinless electrons in $p2$ while assuming time-reversal symmetry. Since both of the $A$ and $B$ irrep of $C_2$ are unchanged by the addition of time-reversal symmetry, nothing is changed in the analysis, and we see that the symmetry indicator group is still $\mathbb Z_2$. As a time-reversal invariant system cannot have a non-zero Chern number, this leaves us with a little dilemma. To resolve it, we note that, as stressed at the end of \sref{sec:overview}, our notion of band structure only considers the possibility of isolating a set of bands at all high-symmetry momenta. In our $p2$ example, we are only imposing the condition at the four isolated points $\Gamma, X, Y$ and $M$. As such, a material with a nontrivial symmetry indicator of $1 \in \mathbb Z_2$ can feature Dirac points at some general momenta. This is indeed the case \cite{PhysRevX.8.031069}, and the system can be viewed as a would-be Chern insulator which becomes gapless due to the presence of time-reversal symmetry.

\subsection{Inversion symmetry \label{sec:inv}}
As our second example, we consider materials with inversion symmetry. In two-dimensions, the calculation is essentially identical to the $p2$ example we studied in details, and so we again arrive at $X_{\rm BS} = \mathbb Z_2$. However, while projective representations for $C_{2z}$ has to be used in the presence of spin-orbit coupling, the inversion symmetry squares to the identity and so it is still represented by $\pm 1$. As such, we conclude that $X_{\rm BS} = \mathbb Z_2$ for two-dimensional system with only lattice translation and inversion symmetry (layer group $p\bar 1$, no.\ 2). With significant spin-orbit coupling and time-reversal symmetry, this $\mathbb Z_2$ is identical to the Fu-Kane parity criterion \cite{PhysRevB.76.045302}, which is equivalent to the $\mathbb Z_2$ index of two-dimensional TI.

The problem becomes more interesting in three dimensions. As discussed in \cite{Po2017}, one finds that
\begin{equation}\label{eq:SI_inversion}
X_{\rm BS} = \mathbb Z_2^3\times \mathbb Z_4,
\end{equation} 
with or without spin-orbit coupling and/ or time-reversal symmetry. Let us first focus on the case of spin-orbit coupled electrons with time-reversal symmetry. 
In this setup, the Fu-Kane parity criterion \cite{PhysRevB.76.045302} states that all the $\mathbb Z_2$ indices (three weak indices and one strong index) can be inferred from the combinations of the inversion eigenvalues. Indeed, the three $\mathbb Z_2$ factors in equation \eref{eq:SI_inversion} are equivalent to the weak index. However, one finds that the strong index is promoted to a $\mathbb Z_4$ factor, which can be defined as \cite{PhysRevX.8.031070}
\begin{equation}\label{eq:}
\kappa_1 =  \left ( \frac{1}{4} \sum_{\vec k_0 \in {\rm TRIMs}} (n_{\vec k_0}^+ - n_{\vec k_0}^- ) \right) ~~{\rm mod}~~ 4, 
\end{equation} 
where TRIMs denote the set of the eight time-reversal invariant momenta.
While $\kappa_1$ is odd for any strong TIs, phases with $\kappa_1 = 2 ~{\rm mod}~ 4$ is insulating, topological, but not detected by the Fu-Kane criterion. 
As it turns out, it was later discovered to correspond to an inversion-symmetric higher-order TCI \cite{Fang2017} which features helical hinge modes. 
With hindsight, the structure of the symmetry indicators naturally suggests that phases with nontrivial even value of $\kappa_1$ should be modeled as two copies of a strong TI \cite{Fang2017, PhysRevX.8.031070}, for which each of the faces can be gapped out but leaving behind one-dimensional gapless modes at the hinges
\cite{PhysRevLett.108.126807, PhysRevLett.110.046404, PhysRevB.95.165443, Benalcazar61,PhysRevB.96.245115,PhysRevLett.119.246402,PhysRevLett.119.246401, Wieder246,Schindlereaat0346, PhysRevB.97.205136}.
It has also been proposed that elemental bismuth realizes this phase \cite{Schindler:2018aa}. This example also highlights the power of the theory of symmetry indicator: it is capable of detecting topological phases without {\it a priori} knowledge on what the full classification is.

If we instead consider a system without time-reversal symmetry, the same analysis has already been done in \cite{Ari}. While one still finds $X_{\rm BS} = \mathbb Z_2^3 \times \mathbb Z_4$, the interpretation is very different. The $\mathbb Z_2$ factors still correspond to the weak indices, but they now correspond to the parity of the three Chern numbers defined along the three independent directions. The even entry of the $\mathbb Z_4$ factor corresponds to the ``axion insulator'' \cite{PhysRevB.78.195424, PhysRevLett.58.1799}, which is characterized by a nontrivial electromagnetic response  and, in fact, is an early example of a higher order TCI featuring chiral hinge modes 
\cite{PhysRevLett.108.126807, PhysRevLett.110.046404, PhysRevB.95.165443}. As phases with an odd entry in the $\mathbb Z_4$ symmetry indicator can be viewed as ``half'' of the axion insulator, it is inconsistent with an energy gap. Indeed, it was proven that such band structures must feature an odd number of pairs of inversion-related Weyl points \cite{PhysRevB.83.245132,Ari}. Again, as the gap closings happen at general momenta, such Weyl semimetals are consistent with our notion of band structures.

\subsection{Time-reversal symmetric materials}
Having studied two specific examples, we now discuss the general physical interpretation of the symmetry indicators for materials with strong spin-orbit coupling and time-reversal symmetry. These results were obtained by a systematic analysis of either anomalous surface states \cite{PhysRevX.8.031070} or the possible phases from layer construction \cite{QuantitativeMappings}, and then relating the obtained phases to the symmetry indicators.
We remark that these developments were also partly inspired by the parallel developments in the classification of interacting TCIs \cite{PhysRevX.7.011020, PhysRevX.8.011040, Xiong_2018, PhysRevB.96.205106, PhysRevB.97.115153, PhysRevB.99.115116, shiozaki2018generalized}.
A key take-home from these analyses is that all the $X_{\rm BS}$-nontrivial band structures are compatible with a full band gap every where in the Brillouin zone, and so, for this case, a nontrivial symmetry indicator signals a topological (crystalline) insulators. 
In addition, explicit formulas for the symmetry indicators in terms of the multiplicities of the symmetry irreps at high-symmetry momenta, which are closer analogues to the Fu-Kane parity criteria and the rotation-based Chern number formula, are provided in  \cite{QuantitativeMappings,PhysRevX.8.031070}.  In particular, \cite{QuantitativeMappings} provides a detailed tabulation of the explicit formulas of the symmetry indicators for each of the space group.

Here, we provide a terse summary of the key results concerning time-reversal symmetric materials with significant spin-orbit coupling:
\begin{enumerate}
\item In the presence of inversion symmetry, the weak and strong TIs are diagnosable from the Fu-Kane parity criteria. Furthermore, a material which can be formally viewed as two copies of a strong TI, say obtained by a double band inversion at one of the TRIM, will be a TCI. The surface signature, however, will generally depend on the additional symmetries that are present. In particular, it could have surface Dirac cones or helical hinge modes, depending on whether or not there exists a two-dimensional surface which respects the protecting symmetry of the TCI.
\item When a mirror symmetry is present, one can define the $\mathbb Z$-valued mirror Chern numbers \cite{PhysRevB.78.045426}. When, in addition, a $C_n$ rotation symmetry is present, the mirror Chern number can be diagnosed modulo $n$.
\item For systems without inversion symmetry but with the improper four-fold rotation (combination of inversion symmetry with $C_4$), $\bar 4$, there is a $\mathbb Z_2$-valued index which detects the strong TI. This index is denoted by $\kappa_4$ in \cite{PhysRevX.8.031070}.
\end{enumerate}

For materials where spin-orbit coupling is negligible, however, one should use the spinless symmetry indicators. In stark contrast to the spin-orbit coupled case, it was found that any band structure with a nontrivial symmetry indicator will have stable nodal lines or points at some general momenta. The results are discussed in details in \cite{PhysRevX.8.031069}.

\subsection{Magnetically ordered materials \label{sec:magnetic}}
In our treatment so far, we have always assumed the spatial symmetries to be described by one of the 230 three dimensional space groups (which include the two-dimensional counterparts), and that time-reversal symmetry is either present or absent. In a magnetically ordered material, however, the magnetic order may preserve certain combinations of spatial and time-reversal symmetry. For instance, in the Neel order, oppositely aligned spins could be related by a translation followed by the time-reversal operation. The symmetries of a magnetic crystal are described by one of the 1,651 magnetic space groups. Among them, $230\times 2 = 460$ are simply the original space groups and their combination with time-reversal symmetry, and so are already covered by our treatment so far. The remaining 1,191 magnetic space groups feature nontrivial magnetic symmetries, and one has to adapt the theory of symmetry indicator to cover these cases. 

The symmetry indicator groups for all magnetic space groups have been exhaustively computed in \cite{Watanabeeaat8685}. An important simplification comes again from the finiteness of the symmetry indicator group, which can be proven directly given the corresponding finiteness for space groups. This allows one to bypass the analysis of compatibility relations, and instead focus on the much simpler problem of constructing atomic insulator representation vectors, which is not any more difficult in the presence of magnetic symmetries. The symmetry indicator group can then be obtained following the discussion in \sref{sec:SNF}. The band topology of weakly correlated materials with frozen magnetic orders can be diagnosed using the same methods described in \sref{sec:TM}. 
However, unlike the time-reversal symmetric cases, the physical interpretation of the symmetry indicators in the magnetic case remains an open problem, with the existing results only covering certain specific cases \cite{PhysRevB.83.245132,Ari,PhysRevB.86.115112, Watanabeeaat8685, PhysRevB.100.165202}.
In particular, some nontrivial indicators correspond to topological nodal (semi-)metal whereas some others correspond to TIs and TCIs, as was highlighted in \sref{sec:inv} for materials with only lattice translation and inversion symmetry \cite{PhysRevB.83.245132,Ari}.

\section{Discussions}
The study of symmetry representations has always been an integral part of the electronic band theory. The theory of symmetry indicators can be viewed as a modernized approach which borrows important insights from the K-theoretical classifications of TIs and TCIs to simplify the analysis. In particular, due to the finiteness of the symmetry indicator groups, we have established that a  comprehensive analysis of the atomic insulators allows one to bypass the conventional analysis on compatibility relations, and perform an efficient diagnosis of the topological properties of materials.

As we have noted in \sref{sec:AI}, a systematic approach to study the equivalence and relations between atomic insulators was instigated by  \cite{PhysRevLett.45.1025}, which culminated in the theory of band representations (BRs)  \cite{PhysRevLett.45.1025,PhysRevB.23.2824,PhysRevLett.61.1005, Bacry, MichelZak}. 
It should be emphasized that the notion of BRs is built upon the specification of symmetry data in the {\it real space}, i.e., the specification of the Wyckoff positions and the irreps of the site symmetry group. In our terminology, a BR can be viewed as a full specification of an atomic insulator.
Upon passing to the momentum space through a procedure similar to what we simply refer to as a ``Fourier transform,'' one can obtain the symmetry representations at the high-symmetry momenta. However, the specification of an atomic insulator in the real space, as is introduced in the theory of BRs, also contains wave-function information which goes beyond the data encoded in the momentum-space representations.
Indeed, in \cite{PhysRevLett.61.1005} it was recognized that topologically distinct BRs could reduce to the same set of symmetry representations in the momentum space, and for such cases the BRs are differentiated by more refined wave-function properties like the Berry phases. 

This is consistent with the precaution we emphasized: generally, a complete topological band theory requires input beyond just the symmetry data, and so one should not expect to arrive at the full classification by simply analyzing the symmetry representations.
Since the theory of symmetry indicators is designed to focus only on the data available from the momentum-space symmetry representations, our notion of the space $\{ {\rm AI}\}$ does not retain all the topological distinction between the underlying atomic insulators. As an example, the states listed in equation \eref{eq:p2_AI_vecs} all correspond to distinct (elementary) BRs \cite{Bacry}, but we have already seen that their momentum-space symmetry representation vectors are not linearly independent.

In view of the development of topological band theory, it is interesting to ask how the theory of BRs could inform us on the topologically nontrivial states.
This task was undertaken in a series of contemporary works \cite{TQC, PhysRevE.96.023310, PhysRevB.97.035139, PhysRevB.97.035138, PhysRevLett.120.266401}, which proposed a classification of topological materials utilizing the framework of BRs. 
The philosophy behind this scheme is similar to our definition of the symmetry indicator group, namely, any mismatch between the real-space picture of atomic insulators and the momentum-space picture of band insulators signals nontrivial band topology. 
More concretely, one can imagine solving all the arrangement of symmetry representations that are consistent with a band gap according to the compatibility relations\footnote{
The technical approach to solving this problem is different from the one presented in this review. The main difference roots in our view that the symmetry representations are elements of an abelian group, which is natural from a K-theoretic point of view \cite{PhysRevX.7.041069}. This perspective was not introduced in the theory of BRs. Instead, a graph-theoretic approach was introduced in \cite{TQC, PhysRevE.96.023310,PhysRevB.97.035138} for solving the compatibility relations.
}, and then compare them against the momentum-space symmetry representations resulting from BRs.
This provides a binary diagnosis for topological insulators, namely, by consulting a pre-computed database \cite{Bilbao_rep, PhysRevE.96.023310}, one can determine whether or not the representation content of a gapped state is consistent with a BR. If it is inconsistent, it cannot be an atomic insulator.

By definition, an elementary BR can never split into a stack of two atomic insulators. It was found in \cite{TQC} that, when analyzed  in the momentum space, certain elementary BRs do split into two sets of bands separated by a band gap. When this happens, and when the band gap is sustained everywhere in the Brillouin zone (i.e., beyond just the high-symmetry momenta), some nontrivial band topology is ensured. This is true even when the split bands appear compatible with atomic insulators in terms of the momentum-space symmetry representations.
However, as an elementary BR may split into an atomic band together with another one with fragile topology \cite{PhysRevLett.121.126402, PhysRevLett.120.266401, bouhon2018wilson}, some further analysis is required to verify if the resulting state has a topological ground state.

Since symmetry data alone cannot lead to the full classification of topological materials, which generally requires wave-function information, K theory  \cite{PhysRevLett.95.016405, Kitaev, Freed2013, PhysRevB.95.235425, PhysRevX.7.041069, shiozaki2018atiyahhirzebruch} and related approaches originating from the study of interacting symmetry-protected topological orders \cite{PhysRevX.7.011020, PhysRevX.8.011040, Xiong_2018, PhysRevB.96.205106, PhysRevB.97.115153, PhysRevB.99.115116, shiozaki2018generalized} remain as the most promising avenues when one is interested in the full classification of stable (i.e., non-fragile) topological phases. Nevertheless, symmetry-based approaches, including the theory of symmetry indicators, could also lend insights into the classification problem. In particular, it can be readily computed without the overhead of the complete K-theoretic approach, and at the same time constrains the possible K-theory results through well-defined relations between the classification group and the symmetry indicator group. In addition, the symmetry-based approach helps clarify which of the K-theoretic distinctions are ultimately concerned with the mutual topology between atomic insulators; such distinctions play a different role compared to the more conventional invariants in the quest of understanding topologically nontrivial materials.  

The theory of symmetry indicators was designed to aid the practical diagnosis of topological materials and to speed up materials discovery. 
This paradigm has already produced fruitful outcomes: comprehensive database searches for topological materials using the symmetry indicators have led to thousands of candidate materials \cite{Tang2019_NP, Zhang2019, Vergniory2019, Tang2019, Tangeaau8725}. As discussed in \sref{sec:TM}, the method is capable of discovering topological (semi-)metals, TIs, and TCIs in one shot. In fact, it is guaranteed that, up to the accuracy of the {\it ab initio} calculations involved in determining the momentum-space symmetry representations, the scheme exposes all the (non-fragile) topological materials that are diagnosable from symmetry data alone. However, we caution that there are three main caveats behind such large-scale materials discovery. First, the materials prediction relies crucially on the crystal structure as input, and one has to further examine the accuracy of these inputs if they come from previous experimental characterization of synthesized samples, or to investigate into the structural stability of the compound if they were theoretically proposed \cite{Zunger2019BewareOP}. Second, the symmetry analysis does not inform on the detailed energetics of the material, and a potential candidate for, say, a TI maybe highly metallic in reality due to the dispersion of the energy bands. Third, as we have emphasized multiple times, in general the symmetry indicators do not detect all possible forms of band topology, and so certain topological materials would not be uncovered by this approach.

Lastly, we note that the theory of symmetry indicators has also been extended to other symmetry setting. As briefly reviewed in \sref{sec:magnetic}, the symmetry indicator groups appropriate for magnetic materials have already been exhaustively computed in \cite{Watanabeeaat8685}, although a comprehensive analysis of their relation to the physical phases of topological materials remains an open question. Another important generalization is to extend the theory to cover the other ten-fold way classes \cite{PhysRevB.55.1142}. This was initiated in \cite{PhysRevB.98.115150}, and the case most relevant for topological superconductors was discussed in \cite{1811.08712}, which also introduced a notion of weak pairing assumption that is particularly suitable for the diagnosis of topological superconductors using the normal-state symmetry representations. 

While particle-hole symmetry was considered in analyzing the physical meaning of the symmetry indicators in  \cite{PhysRevB.98.115150,1811.08712}, it was not incorporated into the definition of trivial states in these early studies. Very recently, the appropriate generalization required was proposed in \cite{Skurativska2019, 1907.13632, ono2019refined, geier2019symmetrybased}, and the symmetry indicators for topological superconductors have been worked out.
Aside from the enrichment of symmetry settings coming from the different possible pairing symmetries, a main challenge in extending the method of symmetry indicators into the context of superconductors was the lack of a clear notion of the trivial limits. This is exemplified by the case of the one-dimensional topological superconductor with zero-energy Majorana edge modes \cite{Kitaev_2001} and its higher-dimensional analogs (dubbed higher-order topological superconductors \cite{PhysRevLett.119.246401, PhysRevB.97.205136, PhysRevB.98.165144,1904.06361,Ahn2019, PhysRevLett.122.236401, PhysRevB.100.054513, PhysRevLett.123.177001}). These phases have the special properties that if one interprets the single-particle Hamiltonian as one describing an insulator instead of a superconductor, then the system becomes an (obstructed) atomic insulator. To properly diagnose the topological nature of such superconductors, one has to incorporate a comparison with the physical Hilbert space in the definition of the trivial limits. This has been achieved systematically in Refs.\ \cite{Skurativska2019, 1907.13632}. With these proper definitions of the trivial limits, the symmetry indicator groups can be computed using similar methods as those discussed in \sref{sec:SI}, and in particular one can prove directly that the indicators groups are again always finite \cite{ono2019refined}. There are, however, two interesting observations when the theory is applied to the study of topological superconductors \cite{ono2019refined}. First, although the refined notion of trivial limit was introduced mainly for the proper diagnosis of phases with zero-dimensional Majorana modes, it was found that certain two-dimensional topological superconductors with helical edge modes also become detectable thanks to the refinement. Second, it was found that a nontrivial symmetry indicator could indicate a gapped or a gapless phase depending on the additional crystalline symmetries that are present. This suggests that an analysis using only a single spatial symmetry (say a rotation) may not reveal the true identity of the phase, and a more holistic approach taking into account all the crystalline symmetries in the system, like the theory of symmetry indicator, is generally more preferable. 
These results may guide the discovery of topological (crystalline) superconductors in the near future.

\ack
I thank Ashvin Vishwanath and Haruki Watanabe for collaboration on the development of the theory of symmetry indicators, and for helpful comments on the manuscript. I also thank 
Ru Chen, Eslam Khalaf, Jeffrey Neaton, Seishiro Ono, Feng Tang, Xiangang Wan and Michael Zaletel for collaborations on related topics. This work is supported by a Pappalardo Fellowship at MIT and a Croucher Foundation Fellowship.\\

\bibliographystyle{unsrt}
\bibliography{ref}
\end{document}